\input epsf.sty
\documentstyle{mn}

\def\ltsim{\raise 2pt \hbox {$<$} \kern-1.1em \lower 4pt \hbox {$\sim$}}
\def\ltapprox{\raise 2pt \hbox {$<$} \kern-1.1em \lower 5pt \hbox {$\approx$}}
\def\gtsim{\raise 2pt \hbox {$>$} \kern-1.1em \lower 4pt \hbox {$\sim$}}
\def\gtapprox{\raise 2pt \hbox {$>$} \kern-1.1em \lower 5pt \hbox {$\approx$}}

\title[The A3558 Cluster Complex]{Radio Properties 
of the Shapley Concentration \\
III. Merging Clusters in the A3558 Complex}
 
\author[T. Venturi et al.]
{T. Venturi$^{1,\star}$, S. Bardelli$^2$, R. Morganti$^{1}$, R.W. Hunstead$^3$\\
$^1$ Istituto di Radioastronomia, CNR,
via Gobetti 101, I-40129 Bologna, Italy \\
$^2$ Osservatorio Astronomico di Bologna, 
via Ranzani 1, I-40126 Bologna, Italy \\
$^3$ School of Physics, University of Sydney, NSW 2006, Australia \\
$\star$ E-mail: tventuri@ira.bo.cnr.it \\
}

\date{Received XX; Accepted XX}
\pagerange{\pageref{firstpage}--\pageref{lastpage}}
\pubyear{}

\begin{document}

\maketitle
 
\begin{abstract}

We present the results of a 22 cm radio survey carried out with the
Australia Telescope Compact Array (ATCA) covering the
A3558 complex, a chain formed by the
merging ACO clusters A3556-A3558-A3562 and the two groups
SC1327$-$312 and SC1323$-$313,
located in the central region
of the Shapley Concentration. The purpose of our survey is to 
study the effects of cluster mergers on the statistical
properties of radio galaxies and to investigate the connection
between mergers and the presence of radio halos and relic sources.

We found that the radio source counts in the A3558 complex are consistent
with the background source counts. The much higher optical
density compared to the background is not reflected as 
a higher density of radio sources.
Furthermore, we found that no correlation exists between 
the local density and the radio source power, and that steep spectrum
radio galaxies are not segregated in denser optical regions.

\noindent
The radio luminosity function for elliptical and S0 galaxies is
significantly lower than
than for cluster early type galaxies and for
those not selected to be in clusters at radio powers logP$_{1.4}$ 
\gtsim$~$ 22.5, 
implying that the probability of a galaxy becoming a radio source 
above this power limit is lower
in the Shapley Concentration compared with any other environment.
Possible explanations will be presented.

The detection of a head-tail source in the
centre of A3562, coupled with careful inspection of the 20 cm
NRAO VLA Sky Survey (NVSS) and of 36 cm MOST observations,
allowed us to spot two extended sources in the
region between A3562 and SC1329$-$313, i.e. a candidate 
radio halo at the centre of A3562, and low brightness extended emission
around a 14.96 magnitude Shapley galaxy. The relation
between these two extended galaxies and the ongoing group merger 
in this region of the Shapley Concentration are discussed.

\end{abstract}

 
\section{Introduction}

Cluster mergers are the most energetic and common phenomena
in the Universe. They are the natural way of forming rich clusters of
galaxies within cold dark matter scenarios, which imply a bottom-up
hierarchy of structure formation. Cosmological numerical simulations indicate
that the matter flows along preferential directions (defined by filaments
or planar structures),
where the subunits are accelerated to velocities of the order of
$10^{3}$ km s$^{-1}$. In particular, the formation of a rich cluster is
expected at the intersections between filaments or walls.
The merging process generates important perturbations in the 
intracluster medium (ICM) such as shocks, bulk flows and turbulence in the 
hot gas (see for instance Roettiger, Burns \& Loken 1996).
The observational signatures of these events are mainly seen in the X--ray 
band, where they appear as distorted isophotes (Slezak, Durret \&
Gerbal 1996) or as regions of
significantly enhanced temperature outside the cluster centre (Markevitch
1996). 
It has also been assumed that these events must significantly affect 
the emission of the galaxy population. 
For instance, in the optical band merging seems to play a role in 
secondary star formation bursts (Caldwell \& Rose 1997)

However, the most spectacular effect of merging is found at radio
wavelength. Bulk motions of the ICM are invoked to be responsible
for bending the wide-angle tail (WAT) objects (Roettiger, Burns 
\& Loken 1993) and for the formation of the U-shape of narrow-angle tail 
(NAT) sources (Bliton et al. 1998),
otherwise impossible to be explained in terms of ram pressure
for these objects with low peculiar velocities with respect to
the cluster centre. Moreover, the bulk motion of the intracluster
gas may provide the energy for particle reacceleration as well as
additional accretion material for the central engine.

\noindent
Two other classes of merging related radio sources are ``relics''
and ``halos''. Radio halos are cluster wide radio sources, 
typically found in the central
regions of galaxy clusters. They usually exhibit a regular
morphology and their dimensions may exceed the megaparsec.
Relics are preferentially found in peripheral cluster regions,
they are usually characterised by an elongated morphology and 
lack an obvious optical counterpart. 
Both types of radio sources are characterised
by low brightness, steep radio spectra ($\alpha$ \gtsim $~$1, for 
S$\propto \nu^{-\alpha}$) 
and  are located in clusters showing
evidence of a recent merger event (Feretti \& Giovannini, 1996).
It is now accepted that merging 
provides at least a large fraction of the energy required to reaccelerate
the electrons deposited in the intracluster medium by extended
radio galaxies (Blandford \& Eichler 1987, Sarazin 1999).

\noindent
The information on the dynamical state of merging clusters
is crucial in order to understand how the radio properties
of the galaxies and of the clusters are influenced by the merging itself.
However, despite the well established connection between such radio sources
and the merging phenomenon, an extensive multifrequency study
of merging clusters is still missing. 

The cores of rich superclusters are the ideal environments to
study the merging phenomenon, since the high peculiar velocities
induced by the enhanced local density favours cluster-cluster
and cluster-group collisions. 
Among rich superclusters, the Shapley Concentration
stands out for its most extreme properties. It is 
the richest nearby supercluster, with 25 member clusters at a density
contrast $\ge 2$ (Zucca et al. 1993), and shows a percentage of interacting 
clusters a factor of three higher than elsewhere (Raychaudhury et al. 1991).
It is therefore a unique laboratory to follow cluster merging and to 
study its signature in a wide range of astrophysical situations.

\vskip 1 truecm \noindent
\section{The A3558 cluster complex in the Shapley Concentration} 

\medskip 
The A3558 cluster complex is a remarkable chain formed by
the three ACO clusters A3556, A3558 and A3562, located at a 
mean redshift z=0.048 ($\sim$ 14500 km s$^{-1}$)
and spanning $\sim$ 7 h$^{-1}$ Mpc (h = H$_0$/100)
almost perpendicular to the line of sight. Two smaller groups,
SC 1327$-$312 and SC 1329$-$313 are located between A3558 and A3562.
The distribution of the optical galaxies with magnitude b$_J \le$ 19.5
is given in Figure \ref{fig:pointings} (upper panel).
This structure is approximately located at the geometrical centre of 
the Shapley Supercluster and can be considered its core.

A detailed analysis of the velocity distribution, made possible 
by the 714 redshifts available in this region
(Bardelli et al. 1998), has revealed that the whole structure is 
characterised by a large number of subcondensations, confirming
that the region is dynamically very active.
The most important properties of the clusters in the A3558 chain
are summarised in Table 1, where we report the J2000 coordinates,
the Bautz-Morgan type, the mean heliocentric velocity and
the velocity dispersion.

% TABLE 1.
%\setcounter{table}{1}
\begin{table*}
\caption[]{Properties of the clusters in the A3558 chain}
\begin{flushleft}
\begin{tabular}{lllcclrc}
\hline\noalign{\smallskip}
Cluster & RA$_{J2000}$ & DEC$_{J2000}$ & B-M Type & R & $<v>$ & $\sigma_v $ 
& kT \\
        &      &       &          &  & km s$^{-1}$ & km s$^{-1}$ 
& 0.1-2.4 KeV  \\
\\
\noalign{\smallskip}
\hline\noalign{\smallskip}
A3556      & 13 24 06  & $-$31 39 45 &  I   &  0 & 14130$^{+42}_{-74}$ & 
411$^{+76}_{-29}$ & - \\
A3558      & 13 27 55  & $-$31 29 24 &  I   &  2 & 14262$^{+75}_{-82}$ & 
992$^{+85}_{-60}$ & 3.3$^1$ \\
SC1327$-$312 & 13 29 47  & $-$31 36 29 &  - &  - & 14844$^{+105}_{-211}$ & 
691$^{+158}_{-146}$  & - \\
SC1329$-$313 & 13 31 36  & $-$31 48 46 &  - &  - & 14790$^{+114}_{-67}$ &
377$^{+93}_{-82}$    & -  \\
A3562      & 13 33 30  & $-$31 40 00 &  I   &  2 & 14492$^{+225}_{-286}$ & 
913$^{189}_{-96}$ & 3.8$^2$ \\
\noalign{\smallskip}
\hline
\end{tabular}

Notes to Table 1.

$<v>$ and $\sigma_v$ are taken from Bardelli et al. (1998).

The coordinates of the centre for A3556, A3558 and A3562 are taken
from 

Abell, Corwin \& Orowin (1989). 

Those for the SC groups are taken from Bardelli et al. (1998).

$^1$ Bardelli et al. 1996; $^2$ Ettori et al. 1997.

\end{flushleft}
\end{table*}
%------- end of table 1

\noindent
{\bf A3558} is the dominant and most massive cluster in the 
complex, with richness class  
originally estimated as 4 (Abell, Corwin \& Olowin, 1989), but
lowered to 2 by Metcalfe et al. (1994). The overestimate
of the optical richness arose from the inclusion
of galaxies from nearby groups and clusters.    
It is dominated by a cD galaxy 
with b$_J$ = 14.26 and v = 14037 km s$^{-1}$, 
%located at $\alpha_{J2000} = 13^h27^m56.7^s, 
%\delta_{J2000} = -31^{\circ}29^{\prime}45.8^{\prime\prime}$,
offcentered with respect to the cluster geometrical centre by
$\sim$ 50 kpc  (Bardelli et al. 1996). 
The mean velocity of the cluster 
is given in Table 1. The structure analysis 
revealed that substructure in A3558 is significant at \gtsim $~3\sigma$ level,
and the presence of nearby groups in this region may have 
caused the overestimate in its optical richness.
Indication of a complex situation comes also from the cluster
X--ray emission. Bardelli et al. (1996) claimed the existence
of a bridge of hot gas connecting A3558 and SC1327$-$312 and possibly
extending to SC1329$-$313.
The X--ray temperature measured by ASCA, k$T$ = 5.5 keV 
(Markevitch \& Vikhlinin 1997), differs from that measured by
ROSAT (see Table 1),
and this could be interpreted as evidence of the presence of 
various co-existing
components in the intracluster gas, at different temperatures.
Moreover, Markevitch \& Vikhlinin (1997) noted a significant 
enhancement of X--ray temperature in the north-eastern part of this cluster,
corresponding to substructure detected in optical studies
(Bardelli et al. 1998). Finally, a moderate cooling flow of 
25 M$_{\odot}$ yr$^{-1}$ centred on the cD galaxy has been detected 
(Bardelli et al. 1996).

{\bf A3562} is the easternmost cluster in the chain, 
and it is dominated by the central 
b$_J$ = 15.09 magnitude cD galaxy with a very extended 
asymmetric halo and velocity v = 14708 km s$^{-1}$.
Given the low number of redshifts in this region of the complex,
a detailed analysis is not possible. The situation is complicated by
possible contamination
from SC 1329-313 and from a filamentary structure crossing the
chain in the east-west direction in the velocity range 
$ 12000 \le v \le 14000$ km s$^{-1}$ (Bardelli et al 1994).
The cluster is an X--ray emitter, as all clusters in the A3558 complex
(see Table 1).

{\bf A3556} is the westernmost cluster in the chain,
with richness class R=0 
and dominated by a central cD galaxy ($b_J=14.42$, $v=14459$ km s$^{-1}$).
Another cluster galaxy of similar luminosity ($b_J=14.32$, 
$v=14074$ km s$^{-1}$) is located at a projected distance of 
$\sim 13^{\prime}$ from the cluster centre.

\noindent
The projected angular separation between A3558 and A3556 is 
$\sim 50^{\prime}$,
less than two Abell radii (1 R$_A$ = 1.77$^{\prime}$/z $\sim 36^{\prime}$). 
As noted by Metcalfe et al. (1994), this distance is less than the 
turnaround radius of A3558, meaning that the two clusters are interacting.

\par\noindent
The substructure analysis carried out by Bardelli et al. (1998)
shows that the velocity distribution is significantly better
fitted by two gaussians with $\langle v \rangle =14130^{+42}_{-74}$  
km s$^{-1}$
and $\sigma=411_{-29}^{+76}$ km s$^{-1}$ and $\langle v \rangle =
15066^{+58}_{-43}$  km s$^{-1}$ and $\sigma=222_{-59}^{+35}$ km s$^{-1}$  
respectively. 
Bardelli et al. (1998) found that the luminosity function of A3556
is very different from that of the other clusters in the chain, being
rather flat and forming a plateau for magnitudes $b_J$ brighter 
than $\sim$ 16.

\par\noindent
No direct measurement of the X-ray flux density is available for 
A3556. On the basis of the cluster dispersion velocity
Ettori et al. (1997) estimated k$T$=2.1 KeV.

\noindent
A3556 has been surveyed at radio wavelengths by Venturi et al. (1997,
hereinafter Paper I), finding that 
all the galaxies forming the bright plateau in the
optical luminosity function are radio loud.
Two extended radio galaxies were found, i.e. 
a very steep spectrum extended radio source and a wide-angle tail (WAT).
The first, associated 
with a $b_J$ = 15.6 magnitude galaxy, is located in the subgroup with 
v = 15066 km s$^{-1}$ infalling towards A3556, and it has been studied 
in detail in Venturi et al. (1998, hereinafter Paper II), where we concluded
that this source is a possible relic.
The WAT is notable because of its large distance from the
centre of A3556, $\sim 26^{\prime}$ corresponding to $\sim$ 1 Mpc, while
WAT sources are normally found close to the cluster centres.
In order to explain the nature of this source, 
Venturi et al. (1997) proposed the existence of gas at such distance
from A3556, in the far  periphery of the complex. 
The presence of this hot gas was confirmed by an 
analysis of the ROSAT All Sky Survey X--ray data
carried out by Kull \& B\"ohringer (1999) .

The two SC groups are located between A3558 and A3562, 
as shown in Figure \ref{fig:pointings} (upper panel), in
coincidence
with the bridge of X--ray emission connecting the cluster centres
(Breen et al. 1994). 
The structure analysis carried out in Bardelli et al. (1998) showed
that SC1329$-$313 is bimodal at $\sim$ 97\% confidence level.
The velocity derived for the second group is 
$\langle v \rangle = 13348^{+69}_{-83}$ km s$^{-1}$, with dispersion
$\sigma_v = 276^{+70}_{-61}$ km s$^{-1}$.

\medskip
Bardelli et al. (1998) proposed that the A3558 complex is 
the remnant of a cluster-cluster collision seen just after the first
core-core encounter, a scenario which is supported observationally
by the substructures present between A3558 and A3562. In this region,
where the position of the shock is expected, 
they found an enhanced fraction of of blue galaxies. 

Very recently, Hanami et al. (1999), studying the apparent ``blue-shift''
of the iron X--ray lines, concluded that SC1329$-$313 presents
clear signs of ongoing or recent ($< 6\times 10^7$ yrs) merging.

\medskip
In this paper we will present and discuss the results of a radio survey 
of the A3558 complex carried out at 22 cm with the Australia Telescope 
Compact Array (ATCA). In addition we will present simultaneous 13 cm (2.3
GHz) observations for all sources with an optical counterpart on the
Digitised Sky Survey (DSS).

The observations and data reduction are presented in Section 3;
in Section 4 we discuss the 22 cm radio sample and its statistics, 
i.e. source counts and logN-logS diagram; in Section 5 we 
deal with the optical identifications; in Section 6 we present
the properties of the Shapley radio galaxies and the relation with
the merging environment; attention will be devoted to
the extended radio sources in A3562 in Section 7; 
our results are summarised and discussed in Section 8.

\par
We assume a Hubble constant H$_0$ = 100 km s$^{-1}$Mpc$^{-1}$.
At the average redshift of the Shapley Concentration, z=0.05, 
$1^{\prime\prime}$ = 0.67 kpc. 
We will assume $S \propto \nu^{-\alpha}$.

\section{Observations and Data Reduction}

\subsection{The 22 cm ATCA observations}

\begin{figure}
\epsfysize=8.5cm
\epsfxsize=\hsize
\epsfbox{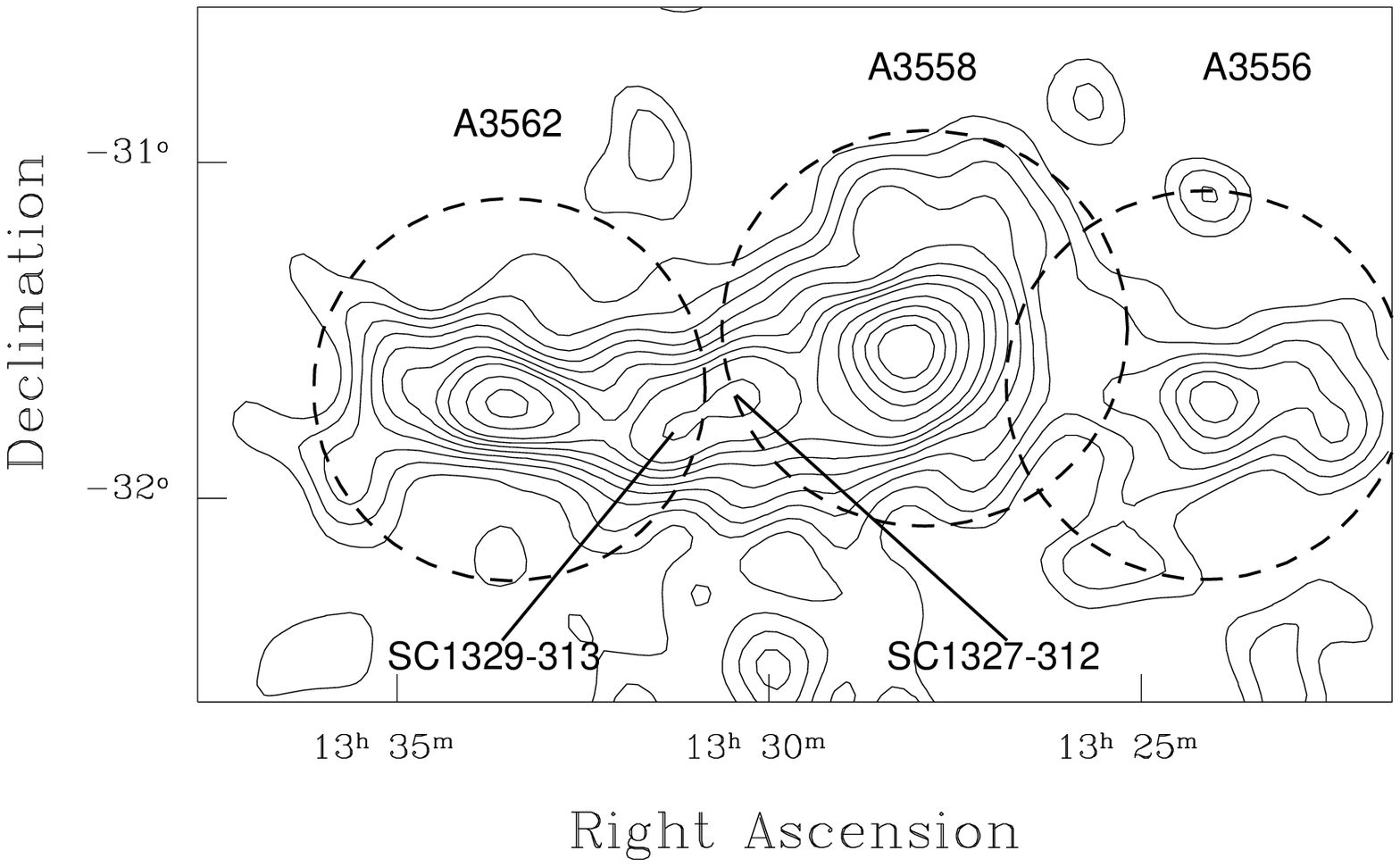}
\epsfysize=8.5cm
\epsfxsize=\hsize
\epsfbox{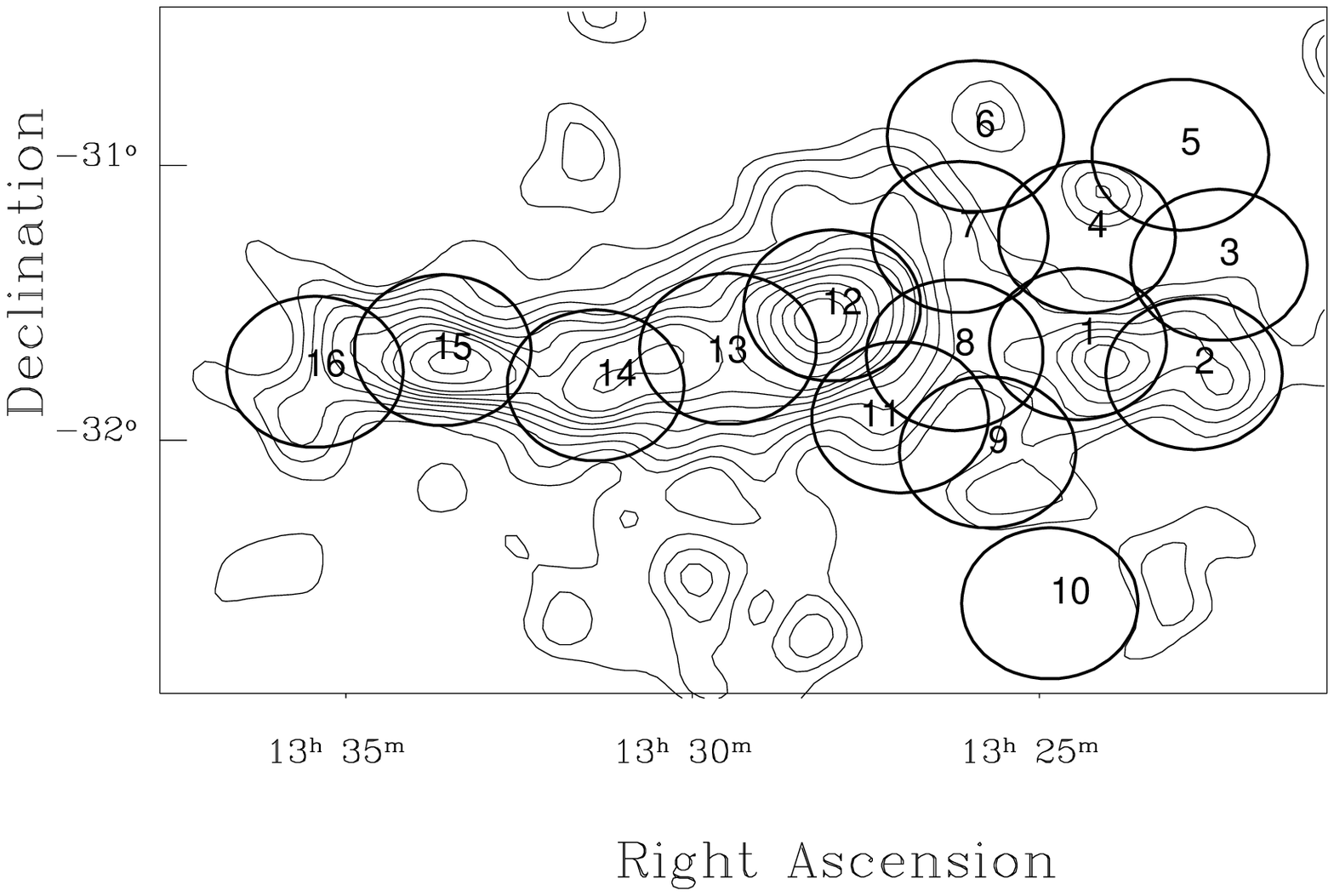}
\caption[]{Isocontours of the galaxy density to b$_J$ = 19.5 in the
A3558 complex. Upper panel: Dashed lines correspond to 1 Abell
radius circles around cluster centres. The positions
of the two poor clusters SC 1327-312 and SC 1329-313 are shown. 
Lower panel:
The centres of the superimposed circles correspond
to the pointing centres of the observations and the circle diameter
is the HPBW of the 22 cm ATCA primary beam. Pointings 1 to 10
are presented in Paper I.}
\label{fig:pointings}
\end{figure}

We observed the A3558 complex with the Australia Telescope Compact
Array at 22 cm ($\nu$ = 1380 MHz), covering the whole region with 16
different pointings and various array configurations.
Table 2 reports the details on the observations and 
Figure \ref{fig:pointings} (lower panel) shows the 
coverage of the A3558 complex with the
observations presented here, superimposed on the optical isodensity
contours. The diameters of the circles are 35$^{\prime}$, corresponding
to the half power primary beam at 22 cm.

%-------------------------------------------------------------------------------
% TABLE 2.
%\setcounter{table}{2}
\begin{table*}
\caption[]{Details on the Observations}
\begin{flushleft}
\begin{tabular}{lllcrccc}
\hline\noalign{\smallskip}
Field & RA$_{J2000}$ & DEC$_{J2000}$ & Array & u-v range & Int. Time & 
rms (22 cm) & rms (13 cm) \\
\#    &              &               &       &    m      &    hr     & 
mJy/b & mJy/b \\
\\
\noalign{\smallskip}
\hline\noalign{\smallskip}
11   & 13 26 59  & $-$31 55 11 & 1.5B + 6C   & 168 - 6000 & 4$\times$ 2 & 
0.1  & 0.09  \\
12   & 13 27 59  & $-$31 30 33 &  6C         & 260 - 6000 & 2           & 
0.2  & 0.16  \\
13   & 13 29 25  & $-$31 40 11 & 1.5B + 6C   & 168 - 6000 & 4$\times$ 2 & 
0.1  & 0.09  \\
14   & 13 31 25  & $-$31 47 57 &  6C         & 260 - 6000 & 2           & 
0.2  & 0.15  \\
15   & 13 33 38  & $-$31 40 17 &  6C         & 260 - 6000 & 2           & 
0.2  & 0.15  \\
16   & 13 35 25  & $-$31 44 59 & 1.5B + 6C   & 168 - 6000 & 4$\times$ 2 & 
0.1  & 0.09  \\
\noalign{\smallskip}
\hline
\end{tabular}
\end{flushleft}
\end{table*}
%------- end of table 2

Pointings \#1 to \#10 were chosen in the A3556 region and the observations
were carried out using the mosaicing facility of the ATCA. For 
details of those observations we refer to Paper I.
Pointings 
\#11, %il nostro field 11
\#13  %il nostro field 12
and \#16 %il nostro field 13
were chosen respectively in order to cover the high density region south of
A3558, the hot gas bridge between A3558 and SC1327$-$312, and 
the outermost periphery of the complex, east of A3562.
In order to cover the whole region of the A3558 chain we also 
reduced and analysed 
archive ATCA data centred on A3558, SC1329$-$313 and on A3562, respectively 
\#12, \#14 and \#15 in Table 1 (see also Reid, Hunstead \& Pierre 1998 for
details of these observations).

% figure 1

All the observations presented in Table 2 and used to cover the
A3558 complex have the resolution of 
$\sim 10^{\prime\prime} \times 5^{\prime\prime}$ in p.a. $\sim 0^{\circ}$.
However, due to the different array configurations and to the
different total time on source, the sensitivity to extended emission 
is not uniform. For example, for fields \#11, \#13 and \#16, 
observed with the configurations 1.5B+6C, the shortest baseline is
$\sim 760 \lambda$, 
%and the largest detectable structure is $\sim 4.5$ arcmin,
while for the remaining fields the shortest spacing is $\sim 1180 \lambda$.
%and the largest detectable structure is $\sim 3$ arcmin.

\noindent
Each observation was carried out with a 128 MHz bandwidth, and the 
correlation of the signal was carried out using 32 channels, each
4 MHz wide, in order to minimise bandwidth smearing effects at large
distance from the pointing centres. 
The data reduction was carried out with the package MIRIAD (Sault,
Teuben \& Wright, 1995), which is particularly suited for the
ATCA observations. Multifrequency synthesis techniques are implemented,
which allow proper gridding of the data in order to reduce 
bandwidth smearing effects.
B1934-638 was used as primary flux calibrator, with an assumed flux density
S$_{22}$ = 14.9 Jy.
The image analysis was carried out with the AIPS package.

\noindent
The noise in the final images varies from field to field (see Table 2). 
For the ten pointings in the A3556 region we assume an average noise of
0.2 mJy/beam (see Paper I). Given the non uniform noise in fields
\#11 to \#16 we chose a detection limit of 1 mJy, corresponding 
to 5$\sigma$ for the fields with the highest noise.
In order to compensate for the sensitivity loss towards the
field edge due to the primary beam attenuation, we corrected
the flux density of the sources in our survey using the analytical
formula of the primary beam attenuation for the Australia
Telescope Compact Array given in Wieringa \& Kesteven (1992).

%
% oppure campione completo a 2.3 mJy includendo sorgenti fino a 22arcmin
% verificare il 22

We estimated that the uncertainty associated with the flux density
measurements is:

$$\Delta S = \sqrt{a^2+(bS)^2}$$

\noindent
where $a$ is the map noise, $b$ is the residual calibration error,
estimated to be of the order of 1\%, and $S$ is the source flux
density.

The radio positional accuracy depends on the
beam size and the source flux density. 
Accordingly we estimate an uncertainty of the order of
$\Delta \alpha = 1^{\prime\prime}$ and $\Delta \delta = 2^{\prime\prime}$
for the weakest sources in the sample.

\subsection{The 13 cm ATCA observations}

13 cm observations were carried out simultaneously with the 22 cm 
observations presented in Sect. 3.1 in each configuration, and the
data reduction was carried out as described in Sect. 3.1. The
adopted flux density for the primary calibrator B1938-638 at this
wavelength is S$_{13}$ = 11.6 Jy. Given the different array configurations
the rms noise in each field varies from 0.09 to 0.16 mJy/beam,
as shown from Table 2.

We used the 13 cm observations to study the spectral index of the
identified radio sources in the A3558 complex field (see Section 5).
The full resolution of the 13 cm images is 
$\sim 6^{\prime\prime} \times 3^{\prime\prime}$. However, for a more
accurate computation of the spectral index (see also Section 6.2) 
we used natural weighted maps, restored with the 22 cm beam.

%
%
% BEGIN TABLE 3
%\setcounter{table}{3}
\begin{table}
\caption[]{Source list and flux density values}
\begin{flushleft}
\begin{tabular}{lllrrrc}
\hline\noalign{\smallskip}
Name       & RA$_{J2000}$   & DEC$_{J2000}$ & S$_{22 cm}$ & R Morph  \\
           &                &               &     mJy     &          \\
\\
\noalign{\smallskip}
\hline\noalign{\smallskip}
J1325$-$3202b & 13 25 25.4 & $-$32 02 47 &   3.0 & unres.  \\
J1325$-$3157  & 13 25 40.7 & $-$31 57 35 & 100.4 & unres.  \\
J1326$-$3140a & 13 26 25.5 & $-$31 40 57 &   1.9 & unres.  \\
J1326$-$3156  & 13 26 48.7 & $-$31 56 43 &   1.4 & unres.  \\
J1326$-$3118  & 13 26 54.6 & $-$31 18 44 &   2.3 & unres.  \\
J1326$-$3123b & 13 26 56.9 & $-$31 23 09 &   1.8 & unres.  \\
J1327$-$3210  & 13 27 10.1 & $-$32 10 31 &   2.0 & unres.  \\
J1327$-$3135  & 13 27 12.7 & $-$31 35 00 &  10.2 & unres.  \\
J1327$-$3202  & 13 27 19.8 & $-$32 02 36 &   5.1 & unres.  \\
J1327$-$3209  & 13 27 24.9 & $-$32 09 39 &  25.6 & res.    \\
J1327$-$3158  & 13 27 26.3 & $-$31 58 15 &   1.2 & unres.  \\
J1327$-$3157  & 13 27 28.3 & $-$31 57 08 &   1.5 & unres.  \\
J1327$-$3123  & 13 27 29.6 & $-$31 23 22 &   2.0 & unres.  \\
J1327$-$3159  & 13 27 38.5 & $-$31 59 01 &   1.3 & unres.  \\
J1327$-$3201a & 13 27 43.1 & $-$32 01 09 &   3.2 & unres.  \\
J1327$-$3201b & 13 27 46.7 & $-$32 01 40 &   1.9 & unres.  \\
J1327$-$3200  & 13 27 46.8 & $-$32 00 30 &   3.4 & unres.  \\
J1327$-$3153  & 13 27 48.7 & $-$31 53 17 &   5.9 & unres.  \\
J1327$-$3152  & 13 27 50.9 & $-$31 52 22 &   1.7 & unres.  \\
J1327$-$3121  & 13 27 50.0 & $-$31 21 01 &  19.7 & D       \\
J1327$-$3129a & 13 27 50.1 & $-$31 29 20 &  43.9 & unres.  \\
J1327$-$3208a & 13 27 51.2 & $-$32 08 30 &   5.9 & unres.  \\
J1327$-$3151  & 13 27 51.6 & $-$31 51 16 &   3.0 & unres.  \\
J1327$-$3208b & 13 27 53.1 & $-$32 08 52 &   4.4 & unres.  \\
J1327$-$3132  & 13 27 54.9 & $-$31 32 18 &   5.7 & unres.  \\
J1327$-$3129b & 13 27 56.8 & $-$31 29 43 &   6.2 & unres.  \\
J1328$-$3133  & 13 28 00.9 & $-$31 33 09 &   1.1 & unres.  \\
J1328$-$3145  & 13 28 02.6 & $-$31 45 20 &  22.9 & unres.  \\
J1328$-$3127  & 13 28 03.1 & $-$31 27 43 &   3.6 & unres.  \\
J1328$-$3129  & 13 28 10.1 & $-$31 29 19 &  10.7 & unres.  \\
J1328$-$3204  & 13 28 11.3 & $-$32 04 01 &  14.6 & unres.  \\
J1328$-$3206  & 13 28 11.8 & $-$32 06 38 &   3.2 & unres.  \\
J1328$-$3148  & 13 28 16.4 & $-$31 48 20 &  64.7 & unres.  \\
J1328$-$3210  & 13 28 20.7 & $-$32 10 56 &   5.1 & unres.  \\
J1328$-$3134  & 13 28 21.2 & $-$31 34 39 &   1.1 & unres.  \\
J1328$-$3111a & 13 28 25.5 & $-$31 11 20 &   2.5 & unres.  \\
J1328$-$3119  & 13 28 29.3 & $-$31 19 30 &  66.3 & D       \\
J1328$-$3135  & 13 28 31.4 & $-$31 35 03 & 119.9 & unres.  \\
J1328$-$3111b & 13 28 34.8 & $-$31 11 55 &   2.6 & unres.  \\
J1328$-$3124  & 13 28 36.2 & $-$31 24 07 &   1.6 & unres.  \\
J1328$-$3139  & 13 28 36.1 & $-$31 39 42 &   2.8 & unres.  \\
J1328$-$3146  & 13 28 44.2 & $-$31 46 27 &   2.7 & unres.  \\
J1328$-$3157  & 13 28 49.7 & $-$31 57 02 &  13.9 & unres.  \\
J1328$-$3209  & 13 28 52.6 & $-$32 09 45 &  13.1 & unres.  \\
J1328$-$3155  & 13 28 56.5 & $-$31 55 24 &   2.5 & unres.  \\
J1328$-$3151  & 13 28 57.6 & $-$31 51 36 &   2.0 & unres.  \\
J1328$-$3144  & 13 28 59.4 & $-$31 44 15 &   2.2 & unres.  \\
J1329$-$3126  & 13 29 00.5 & $-$31 26 45 &  88.1 & FRII    \\
J1329$-$3131  & 13 29 04.6 & $-$31 31 09 &  82.5 & unres.  \\
J1329$-$3129a & 13 29 05.6 & $-$31 29 41 &   4.1 & unres.  \\
J1329$-$3113  & 13 29 08.5 & $-$31 13 05 &   8.6 & unres.  \\
J1329$-$3133  & 13 29 13.0 & $-$31 33 22 &  31.5 & unres.  \\
J1329$-$3121  & 13 29 13.2 & $-$31 21 54 &  22.5 & unres.  \\
J1329$-$3139  & 13 29 28.7 & $-$31 39 25 &   2.6 & unres.  \\
J1329$-$3129b & 13 29 29.5 & $-$31 29 46 &  19.3 & unres.  \\
J1329$-$3116a & 13 29 30.9 & $-$31 16 50 &   7.2 & unres.  \\
\noalign{\smallskip}
\hline
\end{tabular}
\end{flushleft}
\end{table}
%
%
%-- TABLE 3 - continued
\setcounter{table}{2}
\begin{table}
\caption[]{ Continued}
\begin{flushleft}
\begin{tabular}{lllrrrc}
\hline\noalign{\smallskip}
Name       & RA$_{J2000}$  & DEC$_{J2000}$ & S$_{22 cm}$ & R Morph  \\
           &               &               &  mJy        &          \\
\\
\noalign{\smallskip}
\hline\noalign{\smallskip}
J1329$-$3116b & 13 29 31.9 & $-$31 16 55 &   4.2 & unres.  \\
J1329$-$3117  & 13 29 33.5 & $-$31 17 00 &   7.3 & unres.  \\
J1329$-$3158  & 13 29 41.8 & $-$31 58 42 &  56.7 & unres.  \\
J1329$-$3154  & 13 29 41.9 & $-$31 54 14 &   4.7 & unres.  \\
J1329$-$3200  & 13 29 43.7 & $-$32 00 42 &  82.8 & unres.  \\
J1329$-$3122  & 13 29 49.3 & $-$31 22 20 &   4.3 & unres.  \\
J1329$-$3123  & 13 29 51.0 & $-$31 23 00 &  17.3 & FRII  \\
J1329$-$3129  & 13 29 55.3 & $-$31 29 40 &   8.9 & unres.  \\
J1330$-$3145  & 13 30 04.8 & $-$31 45 13 &   3.6 & unres.  \\
J1330$-$3144a & 13 30 05.3 & $-$31 44 57 &   3.9 & unres.  \\
J1330$-$3143  & 13 30 05.7 & $-$31 43 50 &   3.5 & unres.  \\
J1330$-$3144b & 13 30 08.8 & $-$31 44 17 &   3.9 & unres.  \\
J1330$-$3122a & 13 30 19.1 & $-$31 22 59 & 421.7 & unres.  \\
J1330$-$3156  & 13 30 26.9 & $-$31 56 51 &   3.1 & unres.  \\
J1330$-$3130  & 13 30 30.9 & $-$31 30 01 &  14.2 & unres.  \\
J1330$-$3138  & 13 30 37.2 & $-$31 38 31 &   2.3 & unres.  \\
J1330$-$3134  & 13 30 38.7 & $-$31 34 56 &   3.1 & unres.  \\
J1330$-$3141  & 13 30 53.5 & $-$31 41 34 &   4.3 & unres.  \\
J1330$-$3204  & 13 30 55.9 & $-$32 04 06 &   9.0 & unres.  \\
J1330$-$3122b & 13 30 59.1 & $-$31 22 54 &   8.7 & unres.  \\
J1331$-$3144  & 13 31 00.8 & $-$31 44 54 &   1.6 & unres. \\
J1331$-$3155  & 13 31 06.5 & $-$31 55 08 &   7.1 & unres. \\
J1331$-$3139a & 13 31 11.3 & $-$31 39 38 &   2.6 & unres. \\
J1331$-$3139b & 13 31 12.1 & $-$31 39 26 &   8.9 & unres. \\
J1331$-$3138a & 13 31 14.4 & $-$31 38 04 &   7.4 & unres. \\
J1331$-$3128  & 13 31 16.8 & $-$31 28 28 &  18.7 &   res. \\
J1331$-$3210  & 13 31 16.9 & $-$32 10 05 &   3.2 & unres. \\
J1331$-$3143  & 13 31 19.7 & $-$31 43 37 &   7.4 &   res. \\
J1331$-$3149  & 13 31 21.4 & $-$31 49 03 &   1.3 & unres. \\
J1331$-$3145  & 13 31 22.9 & $-$31 45 27 &   1.7 & unres. \\
J1331$-$3137  & 13 31 22.9 & $-$31 37 39 &   1.7 & unres. \\
J1331$-$3121  & 13 31 29.8 & $-$31 21 51 &  11.7 & unres. \\
J1331$-$3138b & 13 31 31.6 & $-$31 38 36 &   6.9 & unres. \\
J1331$-$3125  & 13 31 31.6 & $-$31 25 11 &   4.9 & unres. \\
J1331$-$3206  & 13 31 42.7 & $-$32 06 38 & 113.2 &  FRII  \\
J1331$-$3147a & 13 31 43.9 & $-$31 47 39 &   3.1 & unres. \\
J1331$-$3201  & 13 31 48.9 & $-$32 01 14 &   2.0 & unres. \\
J1331$-$3142  & 13 31 51.0 & $-$31 42 48 &   2.4 & unres. \\
J1331$-$3156  & 13 31 53.7 & $-$31 56 09 &   1.7 & unres. \\
J1332$-$3146  & 13 32 03.1 & $-$31 46 47 &   4.7 & unres. \\
J1332$-$3141a & 13 32 04.6 & $-$31 41 34 &   7.9 &   D    \\
J1332$-$3153a & 13 32 16.0 & $-$31 53 02 &   3.3 & unres. \\
J1332$-$3152  & 13 32 17.6 & $-$31 52 49 &  14.6 & unres. \\
J1332$-$3215  & 13 32 17.1 & $-$32 15 35 &  25.1 & unres. \\
J1332$-$3123  & 13 32 27.5 & $-$31 23 55 &  15.0 & unres. \\
J1332$-$3141b & 13 32 31.8 & $-$31 41 54 &   4.8 & unres. \\
J1332$-$3153b & 13 32 33.7 & $-$31 53 14 &   2.4 & unres. \\
J1332$-$3202  & 13 32 40.0 & $-$32 02 03 &  16.4 & unres. \\
J1332$-$3201  & 13 32 42.5 & $-$32 01 43 &   6.1 & unres. \\
J1332$-$3158a & 13 32 44.0 & $-$31 58 18 &   3.1 & unres. \\
J1332$-$3158b & 13 32 45.2 & $-$31 58 29 &  15.4 & unres. \\
J1333$-$3117  & 13 33 05.3 & $-$31 17 48 &   4.4 & unres. \\
J1333$-$3141  & 13 33 31.6 & $-$31 41 01 &  99.0 &   HT   \\
J1333$-$3130  & 13 33 37.4 & $-$31 30 45 &  37.2 & unres. \\
J1333$-$3118  & 13 33 37.9 & $-$31 18 05 &   3.2 & unres. \\
J1333$-$3135  & 13 33 41.4 & $-$31 35 47 &   1.2 & unres. \\
\noalign{\smallskip}
\hline
\end{tabular}
\end{flushleft}
\end{table}
%
%
%------- TABLE 3 - continued
\setcounter{table}{2}
\begin{table}
\caption[]{ Continued}
\begin{flushleft}
\begin{tabular}{lllrrrc}
\hline\noalign{\smallskip}
Name       & RA$_{J2000}$  & DEC$_{J2000}$ & S$_{22 cm}$ & R Morph  \\
           &               &               &     mJy     &          \\
\\
\noalign{\smallskip}
\hline\noalign{\smallskip}
J1333$-$3120  & 13 33 53.1 & $-$31 20 02 &   9.6 & unres. \\
J1334$-$3146  & 13 34 04.0 & $-$31 46 32 &   3.9 & unres. \\
J1334$-$3128  & 13 34 08.3 & $-$31 28 37 &  29.8 &   D    \\
J1334$-$3136  & 13 34 09.1 & $-$31 36 58 &   1.6 & unres. \\
J1334$-$3123  & 13 34 12.6 & $-$31 23 58 &   4.7 & unres. \\
J1334$-$3149  & 13 34 13.9 & $-$31 49 51 &   6.3 & unres. \\
J1334$-$3119  & 13 34 22.2 & $-$31 19 19 &  19.3 &  res.  \\
J1334$-$3139  & 13 34 22.5 & $-$31 39 07 &  16.0 & unres. \\
J1334$-$3137  & 13 34 30.8 & $-$31 37 45 &   1.4 & unres. \\
J1334$-$3141  & 13 34 35.8 & $-$31 41 04 &   2.0 & unres. \\
J1334$-$3132a & 13 34 36.7 & $-$31 32 42 &   3.9 & unres. \\
J1334$-$3132b & 13 34 37.4 & $-$31 32 49 &  14.7 & unres. \\
J1334$-$3132c & 13 34 40.5 & $-$31 32 45 &   2.2 & unres. \\
J1334$-$3125  & 13 34 51.2 & $-$31 25 35 &   2.1 & unres. \\
J1334$-$3151  & 13 34 52.2 & $-$31 51 15 &   2.5 & unres. \\
J1335$-$3139  & 13 35 03.0 & $-$31 39 18 &  15.8 & unres. \\
J1335$-$3130  & 13 35 07.9 & $-$31 30 42 &   1.6 & unres. \\
J1335$-$3118  & 13 35 07.3 & $-$31 18 07 &  12.8 & unres. \\
J1335$-$3133  & 13 35 12.7 & $-$31 33 43 &   1.5 & unres. \\
J1335$-$3124  & 13 35 16.4 & $-$31 24 57 &   8.6 & unres. \\
J1335$-$3125  & 13 35 17.8 & $-$31 25 08 &   4.3 & unres. \\
J1335$-$3126  & 13 35 20.4 & $-$31 26 20 &   3.2 & unres. \\
J1335$-$3143a & 13 35 27.3 & $-$31 43 05 &   1.8 & unres. \\
J1335$-$3153a & 13 35 42.6 & $-$31 53 53 &  14.4 &  FRI   \\
J1335$-$3117  & 13 35 42.6 & $-$31 17 52 &   6.6 & unres. \\
J1335$-$3140  & 13 35 43.5 & $-$31 40 58 &   1.4 & unres. \\
J1335$-$3153b & 13 35 49.5 & $-$31 53 45 &   4.8 & unres. \\
J1335$-$3143b & 13 35 49.1 & $-$31 43 44 &   1.3 & unres. \\
J1335$-$3146  & 13 35 54.3 & $-$31 46 34 &   9.4 & unres. \\
J1335$-$3200  & 13 35 54.6 & $-$32 00 06 &   6.0 & unres. \\
J1336$-$3144  & 13 36 01.3 & $-$31 44 00 &   3.3 & unres. \\
J1336$-$3146a & 13 36 02.8 & $-$31 46 52 &   3.0 & unres. \\
J1336$-$3122a & 13 36 05.4 & $-$31 22 28 &   4.3 & unres. \\
J1336$-$3140  & 13 36 09.6 & $-$31 40 07 &  12.8 & extended\\
J1336$-$3146b & 13 36 17.9 & $-$31 46 44 &   2.0 & unres. \\
J1336$-$3148  & 13 36 18.6 & $-$31 48 30 &   2.3 & unres. \\
J1336$-$3205  & 13 36 28.2 & $-$32 05 45 &  19.3 & unres. \\
J1336$-$3122b & 13 36 45.6 & $-$31 22 26 &  10.2 & unres. \\
J1337$-$3204  & 13 37 17.5 & $-$32 04 30 & 123.1 &   D    \\
J1337$-$3139  & 13 37 46.8 & $-$31 39 59 &  43.2 & unres. \\
\noalign{\smallskip}
\hline
\end{tabular}
\end{flushleft}
\end{table}
%
% end of table 3
%
% begin of table 4
%-------------------------------------------------------------------------------
% TABLE 1.
%\setcounter{table}{4}
\begin{table}
\caption[]{Radio sources fainter than the sample flux density limit}
\begin{flushleft}
\begin{tabular}{lllrrrc}
\hline\noalign{\smallskip}
Name & RA$_{J2000}$ & DEC$_{J2000}$ & S$_{22 cm}$ & R Morph \\
     &              &                &  mJy       & y &     \\
\\
\noalign{\smallskip}
\hline\noalign{\smallskip}
J1325$-$3157  & 13 25 36.5 & $-$31 57 04 &  1.6 &  unres.  \\
J1326$-$3220  & 13 26 02.2 & $-$32 20 11 &  5.4 &  unres.  \\
J1326$-$3200  & 13 26 16.1 & $-$32 00 31 &  1.0 &  unres.  \\
J1326$-$3155  & 13 26 30.4 & $-$31 55 32 &  1.0 &  unres.  \\
J1327$-$3209  & 13 27 00.9 & $-$32 09 22 &  1.1 &  unres.  \\
J1327$-$3135b & 13 27 08.8 & $-$31 35 27 &  1.6 &  unres.  \\
J1334$-$3143  & 13 34 06.6 & $-$31 43 39 &  1.6 &  unres.  \\
J1335$-$3123  & 13 35 08.4 & $-$31 23 24 &  1.7 &  unres.  \\
J1335$-$3134  & 13 35 18.9 & $-$31 34 57 &  1.1 &  unres.  \\
J1336$-$3134  & 13 36 49.7 & $-$31 34 31 &  2.7 &  unres.  \\
J1337$-$3135  & 13 37 05.1 & $-$31 35 35 &  3.2 &  unres.  \\
J1337$-$3145  & 13 37 26.7 & $-$31 45 22 &  4.2 &  unres.  \\
\noalign{\smallskip}
\hline
\end{tabular}
\end{flushleft}
\end{table}
%
% end of table 4
%
\section{The sample of radio sources}

The total number of sources detected at 22 cm above 1 mJy in the six fields
presented in this paper is 151. 
The list is given in Table 3,
where we report respectively their name, position, flux 
density at 22 cm and comments on the radio morphology.
In Table 4 we list the sources detected in the fields with the lowest
rms noise (fields \#11 and \#16 in Table 1), whose flux density 
is S$ < $ 1 mJy (before the primary beam correction) but greater
than 5$\sigma$, which we consider reliable detections.
All flux densities given in Table 3 and 4 are corrected for the primary 
beam attenuation. Integrated flux densities are given for the extended
radio sources.

As clear from Table 3, 
the majority of the radio sources detected in the present
survey is unresolved;
in particular only 15 radio sources are extended, i.e. 10\% of
the total. Nine of the extended radio sources are doubles.
We found 
three classical FRII radio galaxies, one FRI (Fanaroff \& Riley 1974)
and one head-tail source, and the remaining four have asymmetric
radio emission. For double and FR-type radio sources, the position
given in Table 3 is the barycentre of the radio emission, while 
for extended and resolved sources we give the position of the
radio peak.

Adding to the present sample
all radio sources in the A3556 mosaic observations
(see Paper I) above 1 mJy (the same flux density limit), we obtain a total
of 263 radio sources in the whole A3558 complex.
The total number of resolved radio sources is 31,
i.e. $\sim 11$ \% of the total.

\subsection{Radio source counts}

Given the much higher optical density in the A3558 complex
compared to the background, and the cluster merger occurring in this
region, we computed the source counts
for our radio sample for comparison with the background 
counts (Prandoni et al. 1999), in order to study if the dynamical 
properties of this region result in an enhanced number of
radio sources.

Due to the primary beam attenuation, the
sensitivity of the fields we surveyed is not uniform
(see Section 3), so our sample is not complete to the
flux limit of 1 mJy. Therefore 
in order to carry out a statistical analysis we have 
selected a complete subsample, which we will refer to as the ``reduced
sample''. The ``reduced sample'' includes all radio sources published 
in Paper I and in the present paper with S$_{22 cm} \ge 2.0$ mJy within
a distance of 17.5 arcmin from the centre of their field.
At such a distance the primary beam attenuation is reduced by a factor
of two, so sources with flux density S $\ge 2.0$ mJy are
seen as sources with S $\ge 1.0$ mJy before the correction.

\begin{figure}
\epsfysize=8.5cm
\epsfxsize=\hsize
\epsfbox{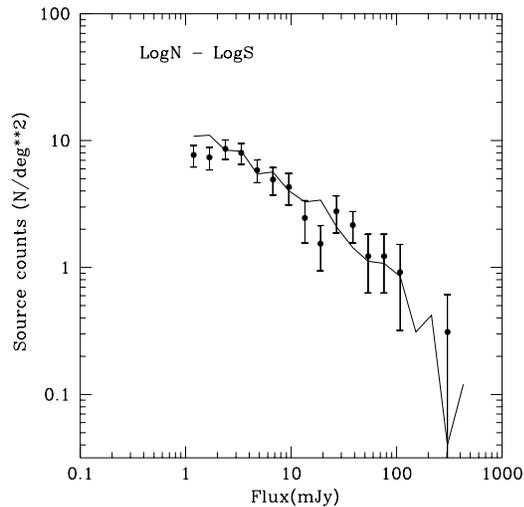}
\caption[]{Differential radio source counts in the A3558 complex.
The filled dots refer to the complete sample presented in this paper. 
The solid line is the logN-logS for the background (Prandoni et al. 1999)}
\label{fig:counts}
\end{figure}

The total number of sources
in the ``reduced sample'' is 145 and the area covered is 3.25 deg$^2$.
The resulting logN - logS in the flux density range 1 - 512 mJy
is reported in Figure \ref{fig:counts}, 
where the differential number of sources in each bin is given in
N deg$^{-2}$ and the errors are poissonian. 
The width of each bin is $\Delta$S = S$\times$log(2.5).
The  reference line in Figure \ref{fig:counts}  
represents the source counts
taken from the ATCA survey of Prandoni et al. (1999), which covers
an area of 25.82 deg$^2$, for a total of 1752 radio sources 
to the flux density limit of 1 mJy at 22 cm. We assume that this survey,
performed with a very similar instrumental configuration is representative
of the source counts in our flux range and can therefore be used as
reference background.

The plot in Figure \ref{fig:counts} clearly shows that the counts in the
two samples have the same shape and normalisation,  
suggesting that the numbers of sources found in the core of 
the Shapley Concentration are consistent with those predicted 
for the background. The consistency in normalisation
between these two distributions also indicates that calibration error
residuals in our sample are negligible.
In Figure \ref{fig:counts} our counts go significantly below the background
for S$\le$ 2 mJy, which independently 
confirms that our sample is incomplete below this limit.
At the 2 mJy limit the expected number of radio sources in each field
from the Prandoni et al. survey (1999) is 12.
The maximum number of observed objects is 19 in field \#13, which
shows only a marginal excess, while the minimum is 8, found in field
\#15, consistent with the expected counts within the errors. 

We have quantitatively estimated the similarity between the source
counts in the A3558 complex and the background counts applying
a KS test to the two distributions, and found that the probability
that they are the same distribution is p=0.996. This result implies
that the major optical overdensity in the core of the Shapley Concentration 
is not reflected into an overdensity of the 
radio sources down to 2 mJy, i.e. logP$_{22}$ (W Hz$^{-1}$) = 21.73.

% figure 2

\section{Optical Identifications}

We searched for optical counterparts of the radio sources
detected in our 22 cm survey using the COSMOS/UKST Southern Sky
Object Catalogue (Yentis et al. 1992), limited to 
$b_J$ = 19.5 (see also Paper I for a discussion on
this limit). Even though the claimed positional accuracy of 
the catalogue is $\sim$ 0.25 arcsec, we adopt a mean positional
error of 1.5 arcsec in order to take into account the error
introduced by the transformation from the plate frame to the sky
(Unewisse et al. 1993, Drinkwater et al. 1995).

In order to test the reliability of the optical counterparts
we used the parameter $R$ defined as:

$$ R^2 = {{\Delta^2_{r-o}} \over {\sigma^2_g + \sigma^2_r}}.$$

\noindent
Here $\Delta^2_{r-o}$ is the offset
between the radio and optical positions, $\sigma^2_g$ is the uncertainty
in the galaxy position and $\sigma^2_r$ is the uncertainty in the
radio position. For point-like radio sources we consider reliable 
identifications those with $R < 3$.

After  cross correlation between our radio sample and the
COSMOS catalogue, we visually inspected the DSS to search for
faint optical counterparts not included in the catalogue.
For such cases we inspected the COSMOS catalogue at fainter limits,
considering also objects not classified as galaxies.
In two cases, J1327$-$3129b and J1333$-$3141, we 
found a bright optical counterpart 
(b$_J$=14.26 and 17.25 respectively) not included in the COSMOS
catalogue. For these objects we adopted the magnitude 
given in Metcalfe et al. (1994).
We found 40 optical counterparts, $\sim$ 26\% of the total.
Furthermore, four radio sources listed in Table 4,
fainter than the 1 mJy limit, have a bright  optical identification.
If we include also the results given in Paper I,
the total number of identified radio sources in the 1 mJy sample of
the A3558 complex, is 69, again $\sim$ 26 \% of the total.

We estimated the completeness and reliability of our sample of identifications
following the method suggested by de Ruiter et al. (1977).
With our limit on $R$ and using formulas (7) and (8) in their paper
we obtained a completeness of 96.3\% and a reliability of
99.6\%.

The list of the identified radio sources for the 
present paper, together with the relevant 
optical information, is given in Table 5. Column 1 reports the 
radio and optical name as given in the COSMOS catalogue, with the exception
of the two galaxies found only in the Metcalfe et al. (1994) list;
in columns 2 and 3 the coordinates (J2000) of the radio source and its
optical counterpart are given; in columns 4 and 5 we report the
flux density of the radio source respectively at 22 cm and 13 cm,
note that in column 4 the b$_J$ magnitude of the counterpart is
also reported; column 6 reports the spectral index
$\alpha_{13}^{22}$; in column 7 we give the monochromatic radio power 
at 22 cm for the radio galaxies with known redshift and the 
absolute magnitude B$_J$ of the optical counterpart; column 8
gives the radio and optical morphologies; column 9 lists the value of 
the parameter R and  the recession velocity.

There is a number of cases where the large optical extent of the galaxy 
and/or the extent of the radio emission lead to $R > 3$. 
In this case we consider the identification reliable if the optical
counterpart falls within the radio isophotes. A note to
Table 5 clarifies these cases. 

\begin{figure}
\epsfysize=8.5cm
\epsfbox{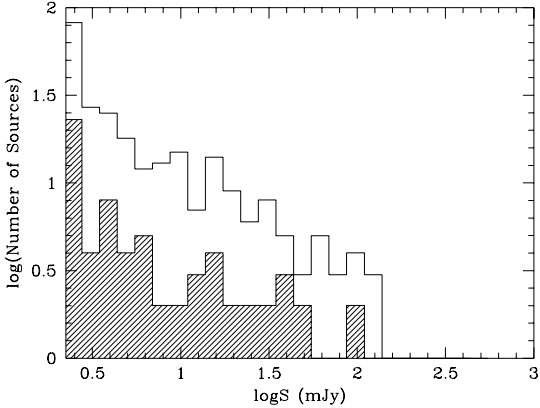}
\caption[]{Distribution of the number of identified radio sources
versus their flux density (hatched bins) and of the total number
of radio sources in the A3558 complex.}
\label{fig:histo}
\end{figure}

In Figure \ref{fig:histo} we show the histogram of the 
number of identified radio
sources as a function of their flux density (shadowed bins). For
comparison the distribution of all radio sources in the A3558 complex
is also shown. The two distributions are remarkably similar,
showing that the optical identification rate is uniform 
over the flux density range of our observations.
%
% Figure 3

\medskip
The redshift information for the optical counterparts is taken from
the spectroscopic sample in Bardelli et al. (1994, 1998).
The sample includes a total of 714 spectra and the global completeness
of the spectroscopic survey at $b_J \le 19.5$ (corresponding to
B$_J \le -16.4$ at the distance of the Shapley Concentration) 
is $\sim 31$\%,
even though it varies considerably from the position with respect
to the cluster centres and the considered magnitude
limit. As shown in Bardelli et al. (1998), all galaxies in
the velocity range 11000 - 17200 km s$^{-1}$ can be considered
part of the complex. 
All the counterparts with pointlike
optical morphology and fainter than $b_J$ = 18.5 without redshift
information have been considered background quasars.
The morphological classification of the optical counterparts given
in Table 5 is done by inspection of the DSS images. No spectral
or photometric information has been taken into account, therefore 
the classification is subject to uncertainties. 

Among the 40 identifications, 19 are located at the redshift
of the Shapley Concentration,
one is a background galaxy (v = 58755 km s$^{-1}$). For six objects
with magnitude $b_J \le 18.0$ there is no redshift information, 
and the remainder are fainter objects.
If we include the results published in Paper I on the A3556 region,
we obtain a total of 28 radio galaxies belonging to the
core of the Shapley Concentration.

With the exception of J1326$-$3118 and J1328$-$3209, which are 
associated with two extended Shapley members likely to be disk galaxies,
all radio emitting galaxies are early type objects.

\begin{figure}
\epsfysize=8.5cm
\epsfxsize=\hsize
\epsfbox{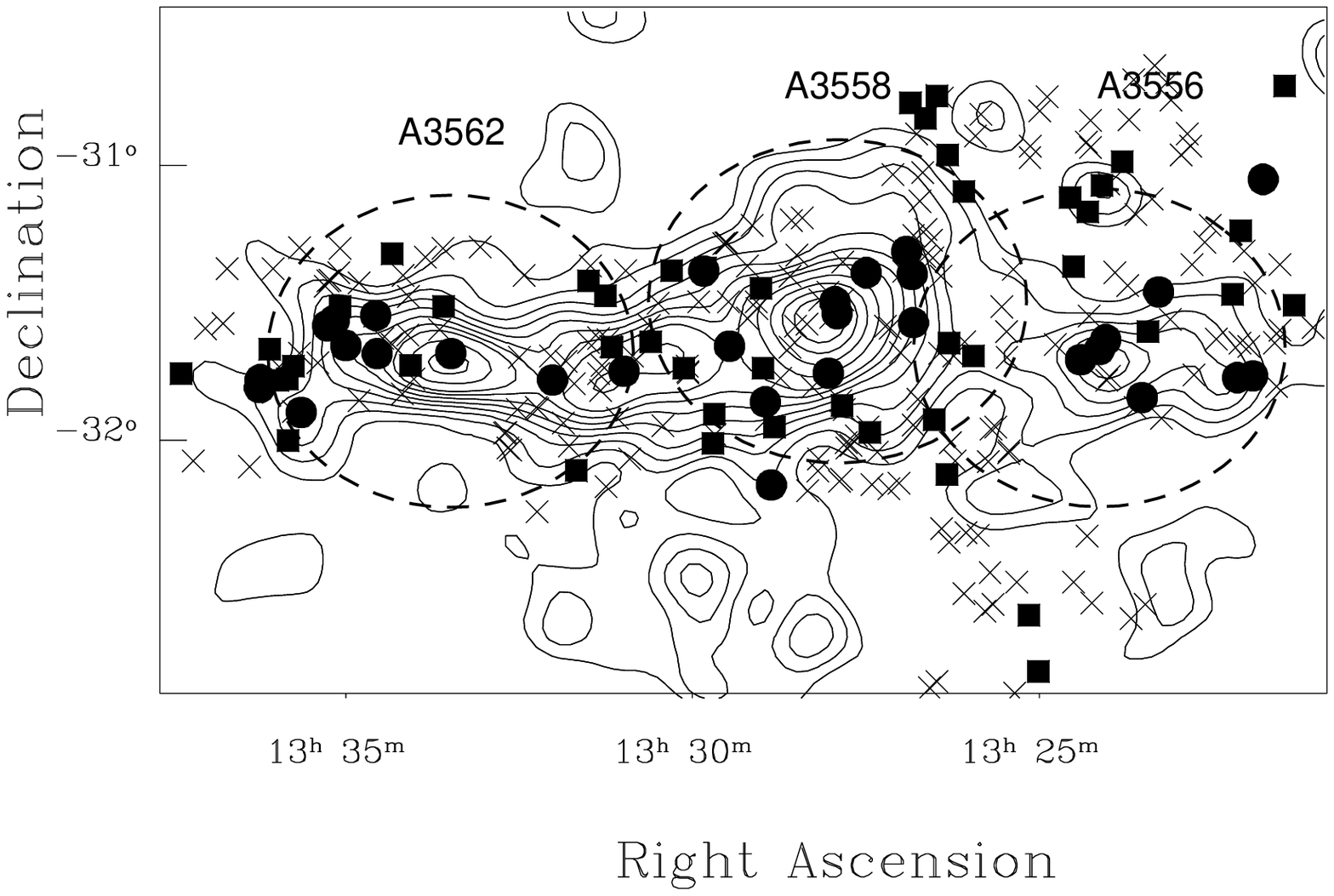}
\caption[]{Location of the radio sources overlaid
on the optical isodensities of the A3558 complex. Filled circles
represent the radio galaxies with measured redshift, filled
squares those without redshift information, crosses stand for
the radio sources without optical counterparts.}
\label{fig:iso}
\end{figure}

\begin{figure}
\epsfysize=8.5cm
\epsfxsize=\hsize
\epsfbox{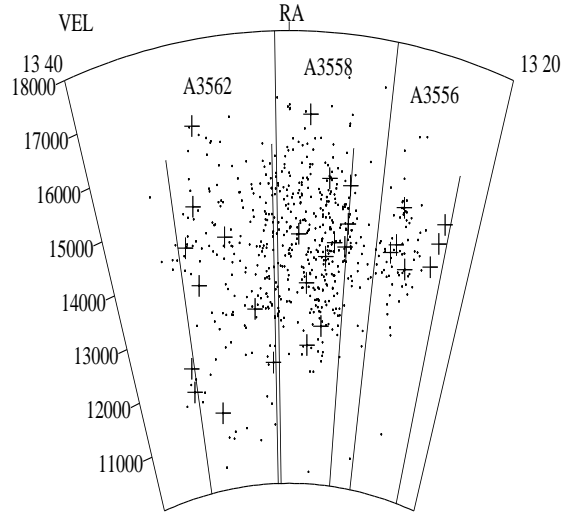}
\epsfysize=8.5cm
\epsfxsize=\hsize
\epsfbox{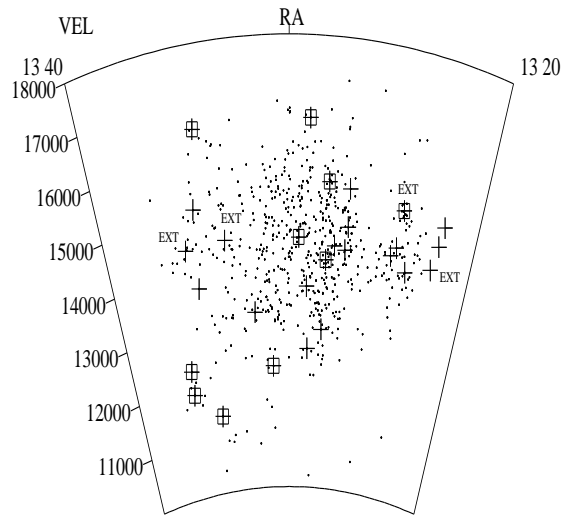}
\caption[]{Upper panel gives the distribution of the A3558 complex 
in the velocity space in the range 10000 - 18000 km s$^{-1}$.
Dots are the optical galaxies, crosses are the Shapley radio galaxies.
Lower panel. Same as upper panel. The 
radio galaxies with 
$\alpha_{13}^{22} \ge 1$ have been marked with a square around the cross
and the extended radio galaxies have also 
been highlighted.}
\label{fig:wedge}
\end{figure}

The location of the
radio galaxies within the complex is well illustrated in Figures \ref{fig:iso}
and \ref{fig:wedge} (upper and lower panel).
Figure \ref{fig:iso} shows the location of the radio sources overlaid
on the optical isodensities of the A3558 complex. 
The distribution of
the radio galaxies belonging to the A3558 complex in 
velocity space is given in Figure \ref{fig:wedge}, where dots are the optical
galaxies and the Shapley radio galaxies are marked with a cross.

%figure 4
%
% figure 5a

%---------------------------------------------------------------------------
% TABLE 4
%\setcounter{table}{5}
\begin{table*}
\caption[]{Optical Identifications}
\begin{flushleft}
\begin{tabular}{lllrrrccc}
\hline\noalign{\smallskip}
Radio Name & RA$_{J2000}$ & DEC$_{J2000}$ & S$_{22}$ & S$_{13}$ & 
$\alpha_{13}^{22}$ & logP$_{22}$ & Radio Type & $R$ \\
   &  &    & mJy & mJy &  & W Hz$^{-1}$ &  & \\
Opt.  Name & RA$_{J2000}$ & DEC$_{J2000}$ & $b_J$  & & & B$_J$ & 
Opt.  Type & v (km s$^{-1})$\\
 \\
\noalign{\smallskip}
\hline\noalign{\smallskip}
J1325$-$3141b& 13 25 56.3 & $-$31 41 33 & 3.8    & $<$ 3.9 &  $> -0.03$ & 
-     & unres. & 0.72 $\star$ \\
           & 13 25 56.2 & $-$31 41 33 & 20.52  &  &  &   &  -     &  -    \\
           &            &           &        &  &  &   &        &       \\
J1326$-$3118 & 13 26 54.6 & $-$31 18 44 & 2.4    & $<$ 4.5 &  $> -1.10$ &
21.76 & unres. & 4.56 $\star$ \\
\#7848     & 13 26 53.9 & $-$31 18 40 & 18.25  &  &  & -17.63 &  S0?  & 14245 \\
           &            &           &        &  &  &   &        &       \\
J1327$-$3123 & 13 27 29.6 & $-$31 23 22 & 2.0    &   1.8   &  0.17  &
21.69 & unres. & 1.79  \\
\#8337     & 13 27 29.5 & $-$31 23 26 & 14.81  &  &  & -21.07  &   E  & 14300 \\
           &            &           &        &  &  &   &        &       \\
J1327$-$3152 & 13 27 50.9 & $-$31 52 22 & 1.7    &   1.1   &  0.76  & 
-    & unres. & 0.98  \\
           & 13 27 50.8 & $-$31 52 24 & 22.04  &  &  &   &   -    &       \\
           &            &           &        &  &  &   &        &       \\
J1327$-$3132 & 13 27 54.9 & $-$31 32 18 & 5.7    &   4.5   &  0.42  &
22.21 & unres. & 0.62  \\
\#8640     & 13 27 54.9 & $-$31 32 19 & 15.13  &  &  & -20.75 &   E   & 15424 \\
           &            &           &        &  &  &   &        &       \\
J1327$-$3129b& 13 27 56.8 & $-$31 29 43 & 6.2    &   1.5   &  2.5   &
22.16 & unres. & 1.12  \\
\#MT 1715  & 13 27 56.9 & $-$31 29 44 & 14.26  &  &  & -21.62 &   E   & 14037 \\
           &            &           &        &  &  &   &        &       \\
J1328$-$3145 & 13 28 02.6 & $-$31 45 20 & 22.9   &  15.7   &  0.68  &
22.65& unres. & 1.15  \\
\#8813     & 13 28 02.5 & $-$31 45 22 & 15.78  &  &  & -20.1  &   E    & 12801 \\
           &            &           &        &  &  &   &        &       \\
J1328$-$3157 & 13 28 49.7 & $-$31 57 02 & 13.9   &  13.3   &  0.08  &
-   & unres. & 1.71  \\
           & 13 28 49.9 & $-$31 57 00 & 17.15  &  &  &   & qso?   &       \\
           &            &           &        &  &  &   &        &       \\
J1328$-$3209 & 13 28 52.6 & $-$32 09 45 & 13.1   &  -      &   -    &
22.38&  res.  & 2.71  \\
\#9411     & 13 28 52.2 & $-$32 09 47 & 16.42  &  &  & -19.46 &  S0?   & 12450 \\
           &            &           &        &  &  &   &        &       \\
J1328$-$3151 & 13 28 57.6 & $-$31 51 36 & 2.0    & $<$ 0.9 & $>$ 1.50 & 
21.81 & unres. & 0.34  \\
\#9550     & 13 28 57.6 & $-$31 51 35 & 17.03  &  &  & -18.85 &   E   & 16532 \\
           &            &           &        &  &  &   &        &       \\
J1328$-$3144 & 13 28 59.4 & $-$31 44 15 & 2.2    & $<$ 5.3 & $>$ -1.56& 
-    &  res.  & 0.87  \\
\#9541     & 13 28 59.2 & $-$31 44 16 & 17.98  &  &  & -17.9  &   E    & 13550 \\
           &            &           &        &  &  &   &        &       \\
J1329$-$3126 & 13 29 00.5 & $-$31 26 45 & 88.1   & 43.0 &  1.27 &
-   &  FRII  & $\star$ \\
           & 13 29 00   & $-$31 26 34 & 18.00  &  &  &   &  qso?  &       \\
           &            &           &        &  &  &   &        &       \\
J1329$-$3139 & 13 29 28.7 & $-$31 39 25 & 2.6    &  1.4 &  1.36 &
21.81 & unres. & 2.18  \\
\#9931     & 13 29 28.4 & $-$31 39 26 & 17.07  &  &  & -18.81 &   E   & 14413 \\
           &            &           &        &  &  &   &        &       \\
J1329$-$3200 & 13 29 43.7 & $-$32 00 42 & 82.8   &  -  &  -  &
-   & unres. & 0.71  \\
           & 13 29 43.6 & $-$32 00 41 & 18.25  &  &  &   & qso?   &       \\
           &            &           &        &  &  &   &        &       \\
J1329$-$3122 & 13 29 49.3 & $-$31 22 20 & 4.3    &  $<$ 3.6 & $>$ 0.30 &
-    & unres. & 0.58  \\
           & 13 29 49.3 & $-$31 22 19 & 18.54  &  &  &   &   -    &       \\
           &            &           &        &  &  &   &        &       \\
J1329$-$3123 & 13 29 51.0 & $-$31 23 00 & 15.5   &  -  &  -  &
23.80&  FRII  & $\star$ \\
\#10178    & 13 29 50.8 & $-$31 22 59 & 19.07  &  &  &   &   E    & 58755 \\
           &            &           &        &  &  &   &        &       \\
J1330$-$3144b& 13 30 08.8 & $-$31 44 17 & 3.2    &  2.3  &  0.56  &
-    & unres. & 0.46  \\
\#10333    & 13 30 08.7 & $-$31 44 17 & 19.53  &  &  &   &   -    &       \\
           &            &           &        &  &  &   &        &       \\
J1330$-$3122a& 13 30 19.1 & $-$31 22 59 & 421.7  & 800.0 & -1.13    &
-  & unres. & 0.35  \\
           & 13 30 19.1 & $-$31 22 59 & 18.46  &  &  &   & qso?   &       \\
           &            &           &        &  &  &   &        &       \\
J1330$-$3138 & 13 30 37.2 & $-$31 38 31 & 2.3    & $<$ 1.6 & $>$ 0.64 &
-    & unres. & 1.14  \\
           & 13 30 37.1 & $-$31 38 33 & 20.68  &  &  &   &   -    &       \\
           &            &           &        &  &  &   &        &       \\
\noalign{\smallskip}
\hline
\end{tabular}
\end{flushleft}
\end{table*}
%------ continuazione tab. 5
\setcounter{table}{4}
\begin{table*}
\caption[]{Optical Identifications. Continued}
\begin{flushleft}
\begin{tabular}{lllrrrccc}
\hline\noalign{\smallskip}
Radio Name & RA$_{J2000}$ & DEC$_{J2000}$ & S$_{22}$ & S$_{13}$ & 
$\alpha_{13}^{22}$ & logP$_{22}$ & Radio Type & $R$ \\
   &  &    & mJy & mJy &  & W Hz$^{-1}$ &  & \\
Opt.  Name & RA$_{J2000}$ & DEC$_{J2000}$ & b$_J$  & & & & 
Opt.  Type & v (km s$^{-1})$\\
 \\
\noalign{\smallskip}
\hline\noalign{\smallskip}
J1331$-$3144 & 13 31 00.8 & $-$31 44 54 & 1.6    & $<$ 0.6 &  $> 1.83$  &
21.45 & unres. & 0.84  \\
\#11000    & 13 31 00.6 & $-$31 44 53 & 16.98  &  &  & -18.9  &   E    & 12141 \\
           &            &           &        &  &  &   &        &       \\
J1331$-$3139a& 13 31 11.3 & $-$31 39 38 & 2.6    & $<$ 0.7 &  $> 2.26$  &
-    & unres. & 1.41  \\
           & 13 31 11.6 & $-$31 39 37 & 17.97  &  &  &   & qso?   &       \\
           &            &           &        &  &  &   &        &       \\
J1331$-$3128 & 13 31 16.8 & $-$31 28 28 & 18.7   & - & - & 
-   &extended& 7.58$\star$ \\
           & 13 31 17.3 & $-$31 28 13 & 17.72  &  &  &   & qso?   &       \\
           &            &           &        &  &  &   &        &       \\
J1331$-$3125 & 13 31 31.6 & $-$31 25 11 & 4.9    & - & - &
-   & unres. & 2.81  \\
           & 13 31 31.5 & $-$31 25 05 & 17.34  &  &  &   & qso?   &       \\
           &            &           &        &  &  &   &        &       \\
J1331$-$3206 & 13 31 42.7 & $-$32 06 38 & 113.1  &  79.3 & 0.63 &
-  &  FRII  & 3.05$\star$ \\
           & 13 31 42.8 & $-$32 06 31 & 17.20  &  &  &   &  qso?  &       \\
           &            &           &        &  &  &   &        &       \\
J1332$-$3146 & 13 32 03.1 & $-$31 46 47 & 4.7    &  3.0  & 0.81 & 
21.98 & unres. & 1.17  \\
\#11744    & 13 32 03.0 & $-$31 46 49 & 14.96  &  &  & -20.92  &   E   & 13107 \\
           &            &           &        &  &  &   &        &       \\
J1333$-$3141 & 13 33 31.6 & $-$31 41 01 & 99.0   & 61.3 & 0.85 &
23.39&  HT    & 0.44  \\
\#MT 4108  & 13 33 31.5 & $-$31 41 00 & 17.25  &  &  & -18.63 &   E    & 14438 \\
           &            &           &        &  &  &   &        &       \\
J1333$-$3130 & 13 33 37.4 & $-$31 30 45 & 37.2   & 19.5 & 1.15 & 
-   & unres. & 0.89  \\
           & 13 33 37.2 & $-$31 30 46 & 22.42  &  &  &   &   -    &       \\
           &            &           &        &  &  &   &        &       \\
J1334$-$3119 & 13 34 22.2 & $-$31 19 19 & 19.3   & - & - & 
-   &extended& 0.91  \\
           & 13 34 22.3 & $-$31 19 18 & 20.63  &  &  &    &   -    &       \\
           &            &           &        &  &  &    &        &       \\
J1334$-$3141 & 13 34 35.8 & $-$31 41 04 & 2.0    & $<$ 0.7 & $>$ 1.77 &
21.49 & unres. & 1.34  \\
\#13198    & 13 34 35.9 & $-$31 41 07 & 16.48  &  &  & -19.4  &   E    & 11357 \\
           &            &           &        &  &  &    &        &       \\
J1334$-$3132c& 13 34 40.5 & $-$31 32 45 & 3.9    & $<$ 1.6 & $>$ 1.55 &
22.10 &  res.  & 5.88$\star$ \\
\#13281    & 13 34 39.9 & $-$31 32 55 & 17.30  &  &  & -18.58 &   E    & 16490 \\
           &            &           &        &  &  &    &        &       \\
J1335$-$3139 & 13 35 03.0 & $-$31 39 18 & 15.8   & 11.1 & 0.63 &
22.63& unres. & 1.07  \\
\#13503    & 13 35 03.1 & $-$31 39 20 & 15.73  &  &  & -20.15 &   E    & 15077 \\
           &            &           &        &  &  &    &        &       \\
J1335$-$3130 & 13 35 07.9 & $-$31 30 42 & 1.6    & $<$ 1.4 & $>$ 0.28 &
-    & unres. & 0.73  \\
           & 13 35 07.8 & $-$31 30 41 & 16.22  &  &  &    &   -    &       \\
           &            &           &        &  &  &    &        &       \\
J1335$-$3133 & 13 35 12.7 & $-$31 33 43 & 1.5    & 1.3 & 0.28 &
21.52 & unres. & 1.10  \\
\#13629    & 13 35 12.6 & $-$31 33 45 & 16.36  &  &  & -19.52 &   E    & 13673 \\
           &            &           &        &  &  &    &        &       \\
J1335$-$3153a& 13 35 42.6 & $-$31 53 53 & 14.4   & 9.2 & 0.79 & 
22.55&  FRI   & 0.84  \\
\#13815    & 13 35 42.5 & $-$31 53 55 & 16.02  &  &  & -19.86 &   E   & 14385 \\
           &            &           &        &  &  &    &        &       \\
J1335$-$3143b& 13 35 49.1 & $-$31 43 44 & 1.3    & 0.8 & 0.84 &
-    & unres. & 1.11  \\
           & 13 35 49.3 & $-$31 43 46 & 21.80  &  &  &    &   -    &       \\
           &            &           &        &  &  &    &        &       \\
J1335$-$3146 & 13 35 54.3 & $-$31 46 34 & 9.4    &  6.5 & 0.66 &
-    & unres. & 1.26  \\
           & 13 35 54.4 & $-$31 46 37 & 22.05  &  &  &    &   -    &       \\
           &            &           &        &  &  &    &        &       \\
J1335$-$3200 & 13 35 54.6 & $-$32 00 06 & 6.0    & - & - &
-    & unres. & 0.58  \\
           & 13 35 54.6 & $-$32 00 07 & 20.81  &  &  &    &   -    &       \\
           &            &           &        &  &  &    &        &       \\
J1336$-$3140 & 13 36 09.6 & $-$31 40 07 & 12.8   & 7.0 & 1.07 &
-   &extended& 0.74  \\ 
           & 13 36 09.7 & $-$31 40 07 & 20.64  &  &  &    &   -    &       \\
           &            &           &        &  &  &    &        &       \\
\noalign{\smallskip}
\hline
\end{tabular}
\end{flushleft}
\end{table*}
%
%------ continuazione tab. 5
\setcounter{table}{4}
\begin{table*}
\caption[]{Optical Identifications. Continued}
\begin{flushleft}
\begin{tabular}{lllrrrccc}
\hline\noalign{\smallskip}
Radio Name & RA$_{J2000}$ & DEC$_{J2000}$ & S$_{22}$ & S$_{13}$ & 
$\alpha_{13}^{22}$ & logP$_{22}$ & Radio Type & $R$ \\
   &  &    & mJy & mJy &  & W Hz$^{-1}$ &  & \\
Opt.  Name & RA$_{J2000}$ & DEC$_{J2000}$ & b$_J$  & & & & 
Opt.  Type & v (km s$^{-1})$\\
 \\
\noalign{\smallskip}
\hline\noalign{\smallskip}
J1336$-$3146b& 13 36 17.9 & $-$31 46 44 & 2.0    & 1.1 & 1.11 &
21.55 & unres. & 0.66  \\
\#14198    & 13 36 18.0 & $-$31 46 46 & 16.06  &  &  & -19.82 &   E   & 12263 \\
           &            &           &        &  &  &   &        &       \\
J1336$-$3148 & 13 36 18.6 & $-$31 48 30 & 2.3    & $<$ 0.8 & $> 1.85$ &
21.58 & unres. & 0.34  \\
\#14199    & 13 36 18.6 & $-$31 48 31 & 16.73  &  &  & -19.15 &   E   & 11839 \\
           &            &           &        &  &  &   &        &       \\
           &            &           &        &  &  &   &        &       \\
\hline\noalign{\smallskip}
J1326$-$3155 & 13 26 30.4 & $-$31 55 32 & 1.0 & - & - &
-    & unres. & 1.30$\diamond$ \\
           & 13 26 30.4 & $-$31 55 35 & 19.42  &  &  &   &   -    &       \\
           &            &           &        &  &  &   &        &       \\
J1334$-$3143 & 13 34 06.6 & $-$31 43 39 & 1.6    & - & - &
-    & unres. & 0.69$\diamond$ \\
\#MT 4345  & 13 34 06.5 & $-$31 43 38 & 18.48  &  &  &   &   -    &       \\
           &            &           &        &  &  &   &        &       \\
J1335$-$3134 & 13 35 18.9 & $-$31 34 57 & 1.1    & $<$ 0.63 & $>$ 0.99 &
22.64 & unres. & 0.33$\diamond$ \\
\#13630    & 13 35 18.9 & $-$31 34 58 & 18.17  &  &  &   &   E    & 58050 \\
           &            &           &        &  &  &   &        &       \\
J1337$-$3145 & 13 37 26.7 & $-$31 45 22 & 4.2    & - & - &
-    &  res.  & 1.20$\diamond$ \\
           & 13 37 26.6 & $-$31 45 24 & 20.17  &  &  &   &   -    &       \\
           &            &           &        &  &  &   &        &       \\
\noalign{\smallskip}
\hline
\end{tabular}

$\star$ Notes to Table 4.

J1325$-$3141b: the source is included in Paper I, where the optical 
ID was missing.

J1326$-$3118: it is located at the east end of an extended edge-on spiral 
(or disk) galaxy.

J1329$-$3126: the candidate optical counterpart is a faint object misplaced
with respect to 

the barycentre of the radio emission.

J1329$-$3123: the optical identification is located in the barycentre of this
double source.

J1331$-$3128: the radio source is extended and the optical counterpart falls
within the radio 

isophotes but it is not coincident with the peak.

J1331$-$3206: the optical counterpart is located in the barycentre of this
double source.

J1334$-$3132c: the radio source is extended and the optical counterpart falls
within the radio 

isophotes but it is not coincident with the peak.

$\diamond$: these radio sources are reliable detections fainter than the
sample flux density 

limit (see Table 3).

The radio sources without 13 cm flux density lie at beyond
the 13 cm field of view.

\end{flushleft}
\end{table*}
%-----------------------------------------------------------------------

\section{Properties of the Shapley radio galaxies}

\subsection{General comments on the radio galaxies}

Two radio galaxies in the survey presented in this paper, both
belonging to A3562, exhibit extended radio emission. 
In particular, J1333$-$3141 is a head-tail 
source located in the centre of the cluster, at a projected distance of
$\sim 1$ arcmin ($\sim$ 40 kpc) from the dominant cD galaxy,
and J1335$-$3153 is located at the eastern periphery
of the cluster and has a double morphology.
These two radio galaxies will be presented in further 
detail in Section 7.

Considering
that a further two extended galaxies were found in A3556, i.e. the tailed 
J1324$-$3138 (whose nature was discussed in Paper II) and the ``mini''
wide-angle tail J1322$-$3146 (Paper I), the total number of
the extended galaxies in the A3558 complex is four.
The remaining radio galaxies are point-like, or show only marginal
extension. 

As clear from Table 5 and from Paper I, all radio galaxies in the 
A3558 complex are faint, their power falling in the range 
logP$_{22}$ (W Hz$^{-1}$) = 21.45 - 23.39,
typical of FRI objects. The strongest source is the head tail
J1333$-$3141. These values suggest that the dominant
mechanism for the radio emission has a nuclear origin. 
We point out that the detection limit of our survey, i.e. 1 mJy,
corresponds to logP$_{22}$ (W Hz$^{-1}$) = 21.42, assuming an average recession
velocity v = 15000 km s$^{-1}$ for the Shapley Concentration. 
Such power is typical of low luminosity radio galaxies
(ellipticals giving rise to FRIs) and of the strongest
spirals.

\subsection{Correlations with the local optical density}

In order to explore the dynamical environment around our radio galaxies,
we cross correlated our sample with the group list of Bardelli
et al. (1998) obtained from their three dimensional sample.
Two out of the 28 Shapley radio galaxies are located in a 
region not covered by the three dimensional analysis. Among the remaining
26, only five are not found in significant  groups. However three
of the five are part of a group just below the significance threshold
and group could be considered an extension of the group T599
(see Table 2 in Bardelli et al. 1998). We found that the groups 
richest in radio galaxies are T337 (corresponding to the main
component of A3558), T598 and T599, both located at the eastern
periphery of A3562. The numbers of radio galaxies
found in these groups are respectively 3, 3 and 4.
This result reinforces the visual impression from Figure \ref{fig:iso} that
the region eastward of A3562 is particularly active at
radio wavelengths.
Conversely, it seems that groups in the regions of SC1329$-$313 and
SC1327$-$312 are void of radio galaxies, although the statistics are poor. 

\begin{figure}
\epsfysize=8.5cm
\epsfbox{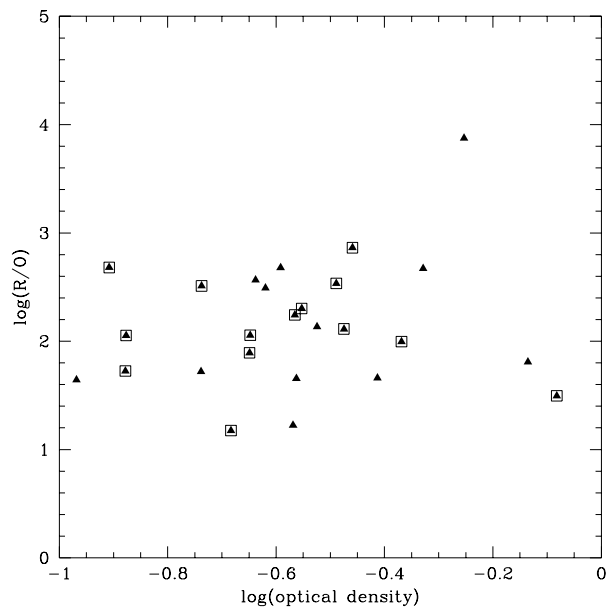}
\caption[]{ Plot of the radio/optical flux of the
Shapley radio galaxies with the local optical density.
The objects enclosed in the empty square are all radio galaxies
with spectral index $\alpha_{13}^{22} \ge$ 1.} 
\label{fig:rorf}
\end{figure}

It has been suggested (Gavazzi \& Jaffe 1986) that there may be 
a correlation between the local optical density and the ratio between
the radio and optical flux (RORF).
In order to check if this behaviour
holds also in the A3558 region, we correlated the local optical density
obtained with the adaptive kernel method (Bardelli et al. 1998)
with the RORF for our radio galaxies. The result of our analysis
is given in Figure \ref{fig:rorf}. No correlation
is found between these two quantities. The radio galaxies in
our sample are distributed in a stripe with constant RORF,
independent of the local galaxy density, indicating that
the power of the radio galaxies is not affected by the local galaxy 
density.
%
% figure 7

It is generally believed (Roland et al. 1985 and references therein) 
that cluster radio sources
are characterised by a steeper spectrum than radio galaxies in other 
environments. This has been explained by invoking confinement of
radiating electrons by the intracluster medium.

In order to see if any segregation effect in the distribution
of the spectral index is present in the Shapley Concentration,
we derived the spectral index $\alpha_{13}^{22}$ for all
radio galaxies in the A3558 complex. For this purpose we used
the full resolution 22 cm images (Section 3.1) and made 
13 cm natural weighted images convolved to the 22 cm HPBW
(Section 3.2). These values are reported in Table 5. 
The average value of $\alpha_{13}^{22}$ for those radio sources
detected both at 22 cm and 13 cm is 0.79, well consistent
with the typical values for radio sources of this class. We point out, 
however, that the presence of six lower limits for $\alpha_{13}^{22}$
(see Table 5), may increase significantly the mean value.
Note that among the four extended Shapley radio galaxies
only the relic source J1324$-$3138 in A3556 has a steep spectrum
($\alpha_{13}^{22} = 1.2$).
We selected all the Shapley radio galaxies with $\alpha_{13}^{22} \ge 1.0$
(including the lower limits)
and highlighted them in Figure \ref{fig:wedge} (lower panel) 
and in Figure \ref{fig:rorf}.
Quite surprisingly, the steep spectrum radio galaxies reside 
preferentially in low density regions at the border of the A3558 
complex. The only exceptions to this behaviour are the cD galaxy
J1327$-$3129b and J1329$-$3139, both  in A3558.
We also find that steep spectrum sources are not segregated
in the density-RORF diagram with respect to the other sources 
(see Figure \ref{fig:rorf}).

\subsection{The radio luminosity function}

The key question we wish to address is whether the ongoing
merging in the A3558 complex has significant influence on
the radio emission of the galaxies. The most direct
method is to compute the radio-optical luminosity
function for the galaxies in this region and compare it
with the mean luminosity functions for cluster galaxies.

In order to compare our radio luminosity function (RLF) for
the Shapley Concentration core to the results obtained by
Ledlow \& Owen (1996, LO96 hereinafter) for a complete sample of 
Abell clusters,
we used only those radio galaxies with an optical counterpart
brighter than b$_J$ = 17.40 ( B$_J = -18.48$) and with
flux density S$_{22} \ge 2.2$ mJy.
The magnitude limit corresponds to the limit M$_R \le -20.5$
in LO96 using the standard conversion b$_J$ = B - 0.2 (B-V) and
the colours for early type galaxies given by 
Fukugita, Shimasaki \& Ichikawa (1995).
At the distance of the Shapley Concentration the 
flux density limit corresponds
to logP$_{22}$ (W Hz${-1}$) = 21.78, the same lower limit as in LO96 after
scaling for the different H$_0$ adopted in their paper. 
These limits in flux density and magnitude reduced the number of 
radio galaxies used to compute the RLF to 17 out of the 28 detected
in the whole A3558 chain (see Section 5). 

The total number of optical galaxies in the A3558 region
covered by our radio survey with b$_J \le$ 17.40 is 216.
To estimate the number of objects actually belonging
to the Shapley Concentration, we corrected this number for
the ratio between the number of the redshifts in the Shapley
velocity range and the total number of redshifts available
in this region, obtaining 203 galaxies. This includes  all morphological
types. For comparison with LO96, however, we need to know the 
fraction of early type galaxies. 
At this magnitude limit, a direct visual
morphological classification is not possible, therefore for 
a reliable estimate of early type galaxies we followed two independent
methods, which take into account respectively {\it (a)} the spectral 
information and {\it (b)} the colour index.

{\it (a)} Assuming that all spectra without emission lines correspond
to early type galaxies, we corrected the total number of Shapley
galaxies for the ratio between non emission line and the total
number of spectra. We obtained 183 objects.

{\it (b)} Assuming that the subsample of Metcalfe et al. (1994) is
representative of our survey, we corrected the total number
of Shapley galaxies for the ratio between galaxies with
B-R $>$ 1.46 and the total number, obtaining 187. The colour
index limit is taken from Fukugita, Shimasaki \& Ichikawa (1995)
and includes ellipticals and S0.

We are aware of the limits of these two indirect methods, 
however the agreement obtained for the estimate of early
type galaxies gives us confidence that it is realistic.
In the computation of the luminosity function we adopted 
185.

\begin{figure}
\epsfysize=8.5cm
\epsfbox{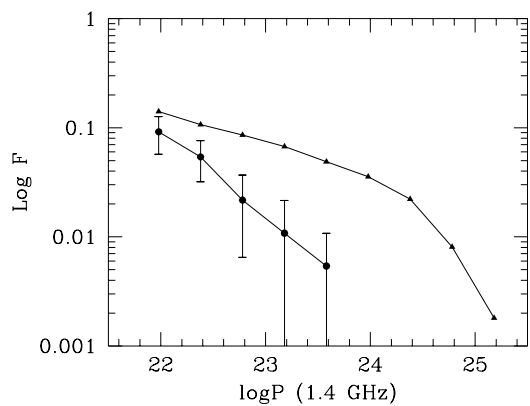}
\caption[]{Radio integral luminosity function, expressed as the
fraction of galaxies emitting with radio power $\ge$ logP,
for the A3558 complex (filled circles) and for cluster galaxies, 
as given in LO96 (filled triangles).}
\label{fig:rlf}
\end{figure}

On the basis
of a simple integration of the LO96 luminosity function, given this number
of optical galaxies, we would expect 26
radio sources against the 17 observed.
The fact that we have fewer radio sources than expected is confirmed
when we compare the differential and integral luminosity function with LO96.
The results of our analysis are given in Table 6, where we give the 
fractional and integral RLF in each power interval. The errors
in each bin are poissonian. 
In Figure \ref{fig:rlf} the integral RLF for the A3558 complex is plotted
together with the cluster RLF derived in LO96.
%
% table 6
%
\begin{table}
\caption[]{Bivariate Luminosity Function}
\begin{flushleft}
\begin{tabular}{ccc}
\hline\noalign{\smallskip}
$\Delta$ logP$_{22}$ & Fractional BLF & Integral BLF \\ 
\\
\noalign{\smallskip}
\hline\noalign{\smallskip}
21.78 - 22.18       & 7/185 & 0.0918 \\
22.18 - 22.58       & 6/185 & 0.0540 \\
22.58 - 22.98       & 2/185 & 0.0216 \\
22.98 - 23.38       & 1/185 & 0.0108 \\
23.38 - 23.78       & 1/185 & 0.0054 \\
\noalign{\smallskip}
\hline
\end{tabular}
\end{flushleft}
\end{table}

It is clear that the two RLFs are significantly 
different, both in shape and
scale, even taking into account the errors and the uncertainties in our
estimate of the number of Shapley E+S0 galaxies. 
The cluster RLF shows a break at logP (W Hz$^{-1}$) $\sim$ 24.4, 
while our RLF has
a decreasing trend without break, and can be described by a single
power law. Applying a KS test to the two distributions we find that
the probability that they are the same is only $\sim 12$\%. 
Even allowing for the errors, the RLF derived for the A3558 complex
is lower than LO96. This result suggests that the probability
of a galaxy in the Shapley Concentration core to become a radio
galaxy is lower than for the comparison sample 
of cluster galaxies from LO96, at least for logP$_{1.4}$ \gtsim$~$ 22.5.
After a statistical comparison with RLFs derived for galaxies not
selected in clusters, LO96 confirmed the results already obtained
by Fanti (1984) and concluded that the radio luminosity function does
not depend on the local galaxy density. Our RLF is therefore different
from those obtained in other environments.

\section{The peculiar radio properties of A3562}

\subsection{The radio galaxies}

As is clear from Table 5 and from Figures \ref{fig:iso} and
\ref{fig:wedge}, nine of the 28
Shapley radio galaxies are located in A3562, the easternmost
cluster in the chain, and most remarkably
seven of them are located at the eastern edge of the cluster (see
also Section 6.2).
Two cluster radio galaxies, J1333$-$3141 and J1335$-$3153,
exhibit extended emission.
The other two extended radio galaxies in the A3558 complex,  
J1322$-$3146 and J1324$-$3138, are located in A3556, at the westernmost
end of the chain, and have already been studied and discussed
in Paper I and Paper II. In this Section we will concentrate on the
radio properties of A3562.

\begin{figure}
\epsfysize=8.5cm
\epsfbox{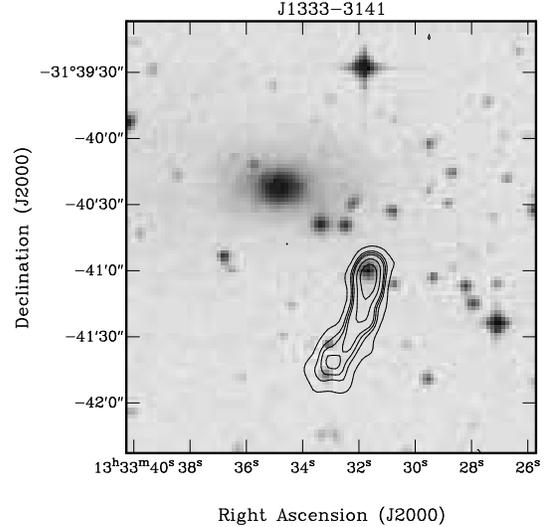}
\caption[]{20 cm full resolution image of the radio source J1333$-$3141
superimposed on the DSS image.
The FWHM of the restoring beam is $10.2 \times 6.5$ arcsec$^2$, 
in p.a. 4.1$^{\circ}$. 
Contours are -0.3, 0.3, 0.75, 1, 1.125, 1.5, 2, 3 mJy/beam.}
\label{fig:ht_20}
\end{figure}

\medskip
The 22  cm radio emission of A3562 is dominated by the head-tail
source {\bf J1333$-$3141} shown in Figure \ref{fig:ht_20},
associated with a 17.25 magnitude elliptical
galaxy (see Table 5), at a projected distance of $\sim 1^{\prime}$
($\sim 40$  kpc) 
from the cluster dominant cD. The bent shape of the tail suggests
motion around the cD galaxy,
possibly on a projected counter-clockwise orbit. 
The velocity difference between the cD galaxy and J1333$-$3141
is $\Delta$v = 110 km s$^{-1}$.
The total
extent of this radio galaxy is $\sim  1^{\prime}$, corresponding
to a projected linear size $\sim$ 40 kpc.

\begin{figure}
\epsfysize=8.5cm
\epsfbox{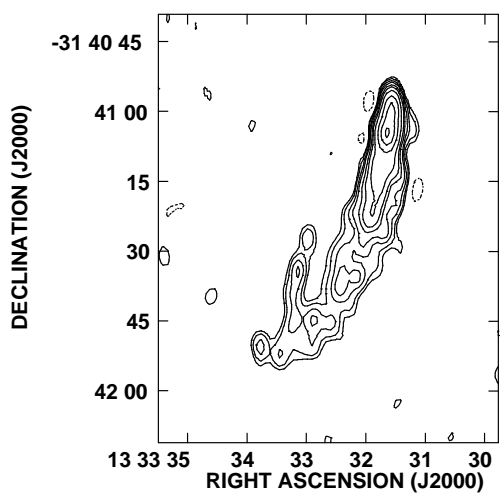}
\caption[]{13 cm full resolution image of the head-tail source J1333$-$3141.
The FWHM of the restoring beam is $5.4 \times 3.2$ arcsec$^2$, 
in p.a. 0.7$^{\circ}$. The peak in the image is 10.5 mJy/beam. 
Contours are -0.35, 0.35, 0.5, 0.5, 0.75, 1, 1.5, 2, 3, 5, 7.5, 10 mJy/beam}
\label{fig:ht_13}
\end{figure}

Our full resolution 22 cm and natural weighted 13 cm images 
give a total spectral index 
$\alpha_{13}^{22} = 0.85$, typical for this type of sources.
Assuming that the peak in the two images corresponds to the
same region, we find that the spectral index is flattest
in the peak, with $\alpha_{13}^{22} = 0.79$, and steepens smoothly
along the tail, increasing to a value $\alpha_{13}^{22} = 1.86$ at
43$^{\prime\prime}$ from the peak. The brightness 
decreases smoothly along the tail, without secondary peaks of emission.
The full resolution
13 cm image given in Figure \ref{fig:ht_13} clearly shows that the 
surface brightness of the tail drops at $\sim$ 25 arcseconds from the core.
This image also suggests that the nuclear region is complex, and that
the peak in the 22 cm full resolution image and in the 13 cm
natural weighted map are possibly the beginning of the bent twin
jets not resolved by our observations rather than
the core of the radio emission.
In order to derive an estimate of the physical parameters in
the tail, we used our value $\alpha_{13}^{22} = 0.85$ for the spectral 
index and we computed the
magnetic field B$_{eq}$ in the source, the minimum non-thermal energy
u$_{min}$ and the minimum non-thermal pressure P$_{nt}$ 
under the hypothesis of equipartition and assuming cilindrical
symmetry. We obtained 
B$_{eq} = 3.0 \times 10^{-6} \mu$G,
P$_{nt} = 0.6 \times 10^{-12}$ dyne cm$^{-2}$,
u$_{min} = 0.9 \times 10^{-12}$ erg cm$^{-3}$.
Such values, which should be considered indicative,
are in the range typical of tailed radio galaxies
in clusters, though at the lower end (Feretti, Perola \& Fanti, 1992).

A multifrequency and multiresolution
study of this source is in progress and the results will be presented
in a future paper.

% figure 7

\begin{figure}
\epsfysize=8.5cm
\epsfbox{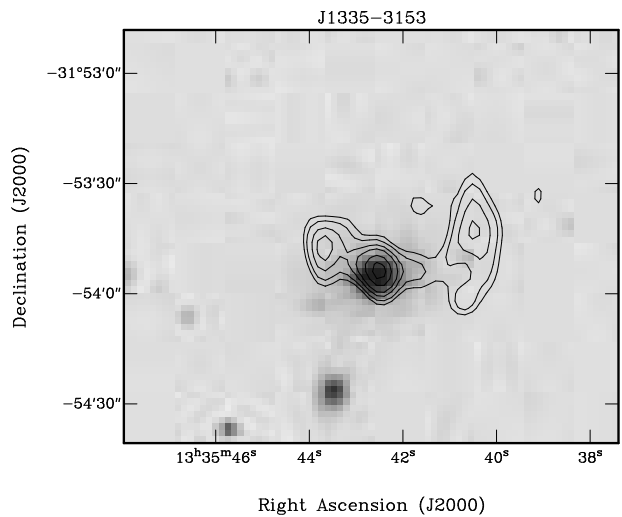}
\caption[]{20 cm full resolution image of the radio source J1335$-$3153
superimposed on the DSS image.
The FWHM of the restoring beam is $10.2 \times 6.5$ arcsec$^2$, 
in p.a. 4.1$^{\circ}$. 
Contours are -0.35, 0.35, 0.5, 0.5, 0.75, 1, 1.5, 2, 3, 5, mJy/beam}
\label{fig:fr1_20}
\end{figure}

{\bf J1335$-$3153}, shown in Figure \ref{fig:fr1_20}, 
is a low luminosity radio source 
associated with a 16.02 magnitude
galaxy with velocity v = 14385 km s$^{-1}$. It is located
at $\sim$ 31.4 arcmin from the centre of A3562 ($\sim 1.26$ Mpc), 
towards the extreme
eastern edge of the cluster. The lobes of this radio galaxy
are not symmetric, the western one being more extended.
From our ATCA
observations we derived a total spectral index
$\alpha_{13}^{22} = 0.79$. The spectrum of the core is flatter,
i.e.   $\alpha_{13}^{22} = 0.53$.
With these values for the spectral index
and assuming equipartition holds in this source we derived
B$_{eq} = 1.9 \times 10^{-6} \mu$G,
P$_{nt} = 0.2 \times 10^{-12}$ dyn cm$^{-2}$ and
u$_{min} = 0.3 \times 10^{-12}$ erg cm$^{-3}$
while for the central component we obtained
B$_{eq} = 3.3 \times 10^{-6} \mu$G,
P$_{nt} = 0.7 \times 10^{-12}$ dyn cm$^{-2}$,
u$_{min} = 1.0 \times 10^{-12}$ erg cm$^{-3}$.

The region where this source is located is at the extreme
eastern edge of A3562, where the X--ray counts fall considerably.
According to Kull \& B\"ohringer (1999), the number of counts in the
bin where this source is located correspond to an X--ray flux density
of the order of $\sim 5 \times 10^{-15}$ erg s$^{-1}$ cm$^{-2}$.
%are of the order of 
%$\sim 5 \times 10^{-4}$ cts arcmin$^{-2}$s$^{-1}$.
A multifrequency and multiresolution study of this source
is being carried out.

\subsection{A candidate radio halo at the centre of A3562 and 
extended radio emission around a peripheral cluster galaxy}

The association between cluster radio halos, diffuse cluster
X--ray emission, the presence of one or more head-tail 
sources near the cluster centre and a recent merging event 
are now widely accepted
(see for instance Feretti \& Giovannini 1996, Feretti et al. 1997, 
Feretti et al. 1990).

The presence of a head-tail radio galaxy located close
to the centre of A3562, the X--ray emission coming from
the centre itself and the ongoing merger in the A3558 complex 
led us to search for extended cluster emission in the centre of A3562
by inspecting the 1.4 GHz NRAO VLA Sky Survey (NVSS, Condon et al. 1998),
whose sensitivity and resolution is well suited for
the detection of extended low surface brightness emission. 

\begin{figure*}
\epsfysize=8.5cm
\epsfbox{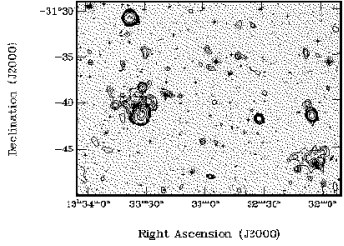}
\caption[]{NVSS 20 cm contours of the Shapley region extending
westward of the centre of A3562, in the direction of the
SC1329$-$313 group, superimposed on the DSS image. 
Contour levels are -1.25, 1, 1.25, 1.5, 2, 2.5, 3, 5 10, 30, 50 mJy/beam.}
\label{fig:halos}
\end{figure*}

\begin{figure}
\epsfysize=8.5cm
\epsfxsize=8.5cm
\epsfbox{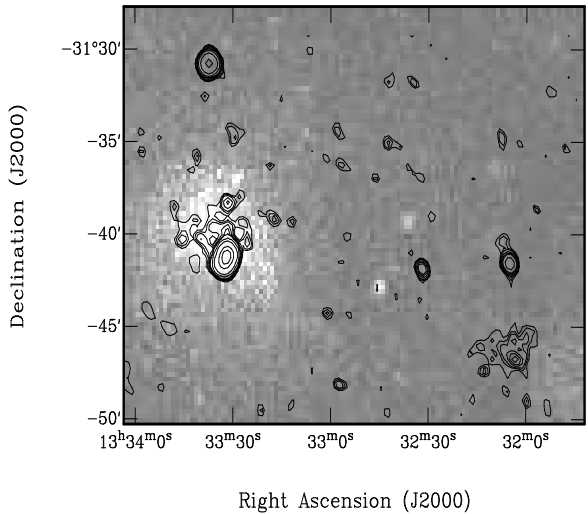}
\caption[]{Same field as Figure 11 with NVSS 20 cm contours
superimposed to a pointed ROSAT PSPC X-ray image.
Radio contours are the same as Figure 11.}
\label{fig:halo_x}
\end{figure}

In Figure \ref{fig:halos} we show the 20 cm (1.4 GHz) NVSS image of the 
field between the centre of A3562 and the group SC1329$-$313
superimposed on the DSS optical frame, and in Figure \ref{fig:halo_x}
the same radio image is superimposed to the X--ray image
taken from the ROSAT All Sky Survey archive.
Both images show a wide field, i.e. $30^{\prime} \times 15^{\prime}$,
which includes the centre of A3562 and the galaxy \#11744 (see below). 
Inspection of Figure \ref{fig:halos} suggests that despite the
presence of discrete sources,  extended emission is
indeed present at the centre of A3562,
eastward of J1333$-$3141 in the direction
of the dominant cD galaxy and superimposed on the X--ray cluster emission
(Figure \ref{fig:halo_x}).

\begin{figure}
\epsfysize=8.5cm
\epsfbox{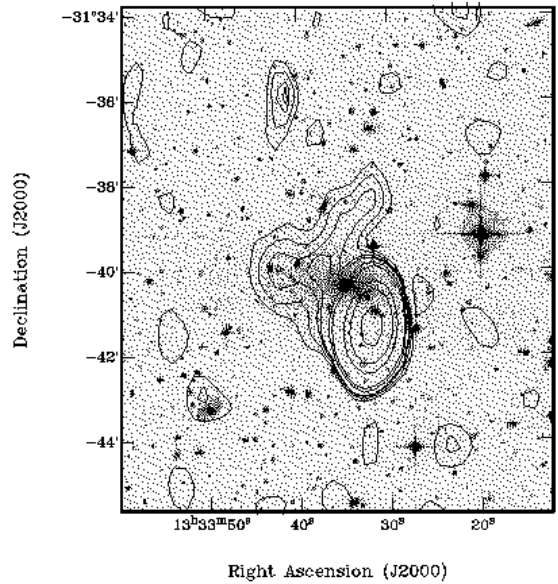}
\caption[]{36 cm MOST contours of the central region in A3562
superimposed on the DSS image. The restoring beam is 81$\times$43
arcsec$^2$, in p.a. 0$^{\circ}$. Contour levels are -2, 2, 3, 4, 4.5, 
7.5, 10, 30, 50, 100 mJy/beam.}
\label{fig:halo_most}
\end{figure}

Given the large HPBW of the NVSS 
($45^{\prime\prime} \times 45^{\prime\prime}$), the head-tail
J1333$-$3141 is only marginally resolved in the NVSS image.
Extended emission is visible also in a 36 cm (843 MHz) image
observed with the Molonglo Observatory Synthesis Telescope
(MOST, Robertson 1991) by Hunstead et al. (in preparation),
which we report in Figure \ref{fig:halo_most}, where the
contours of the 36 cm MOST image are superimposed on the DSS 
optical image. The extension and morphology of the radio
emission from the centre of A3562 at 20 cm (NVSS) and 36 cm
(MOST) are in very good agreement, confirming that this
feature is real.

This emission has an elongated structure, asymmetric with
respect to the location of the head-tail, and from Figures
\ref{fig:halos} and \ref{fig:halo_most} not obviously
connected with J1333$-$3141.
A detailed study on this radio extension and its possible 
connection with J1333$-$3141 is in progress.

% figure 8b

% figure 8a

\begin{figure}
\epsfysize=8.5cm
\epsfbox{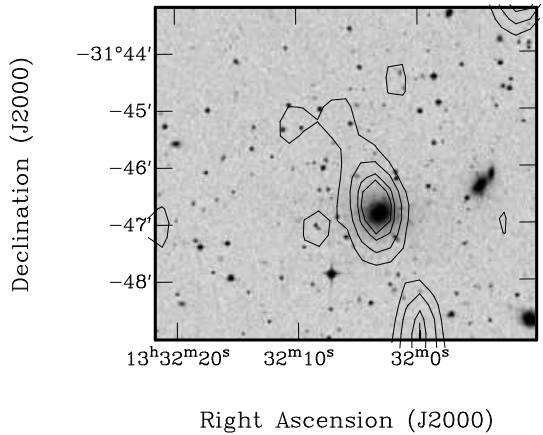}
\caption[]{36 cm MOST contours of the radio galaxy J1332$-$3146
superimposed on the DSS image. The restoring beam is 81$\times$43
arcsec$^2$, in p.a. 0$^{\circ}$. Contour levels are -2, 2, 3, 4, 4.5, 
7.5, 10 mJy/beam.}
\label{fig:minihalo}
\end{figure}

Another remarkable feature of the radio emission in A3562 
evident
from of Figure \ref{fig:halos} and \ref{fig:halo_x} is 
the existence of very low brightness extended emission 
around the radio galaxy J1332$-$3146, detected as a point-like
source in our survey, and identified with the 
14.96 magnitude galaxy \#11744, with recession velocity 
v = 13107 km s$^{-1}$ (see Table 5).
This galaxy is located $\sim$ 21 arcmin ($\sim$ 0.8 Mpc) 
westward of the A3562 centre, 
at a distance of
$\sim$ 6 arcmin ($\sim$ 240 kpc) from the dominant galaxy in SC1329$-$313.
This region of the A3558 complex is 
characterised by the presence of many peculiar and ``disturbed'' galaxies
where a very high fraction of blue galaxies has been recently
found by Bardelli et al. (1998).
The projected angular size of this extended emission is  
$\sim 4^{\prime} \times 2^{\prime}$, corresponding to $\sim 160 \times 80$
kpc, and the flux density derived
from the NVSS image is S$_{20}$ = 21.4 mJy. 
The very low surface brightness of this source, coupled with its
distance from the centre of the field where it was detected, i.e. \#14,
put the extended emission below the sensitivity limit of our
survey. This source was also detected in the MOST 36 cm observations
centred on A3562 (Hunstead et al., in preparation) and the 36 cm 
radio contours are given in Figure \ref{fig:minihalo}. 

The possible radio halo in the centre of A3562 and
the presence of an extended halo around a cluster radio galaxy
are probably related to the merging state of the A3558 complex
and further discussed in Section 8. 

% figure 8c

\section{Discussion and Conclusions}

The aim of this work is to investigate whether and how the cluster merging
process is able to modify the radio properties  of the galaxy population.
From a theoretical point of view, the situation is not clear. Using
simulations, Bekki (1999) found that the merging drives efficient transfer 
of gas to the central regions of galaxies, a good mechanism for feeding
the central engine of AGNs and switching on a starburst. On the other hand,
Fujita et al. (1999) concluded that gas stripping due to ram pressure is 
important in preventing gas supply to the galaxy central regions.     

The ideal environment for studying the merging phenomenon
is the centre of the Shapley Concentration. 
In particular,  evidence
has been accumulated that the A3558 complex is at a late
stage of the merging of clusters of similar mass (Bardelli et al. 1996, 1998). 
Therefore, our approach is to study the radio emission
of galaxies in a place where the existence of a major
merging is established.
This is different from the customary approach, where
cluster interaction is found a posteriori, after detection of
a particular phenomenon.

The results of our ATCA 22 cm survey of the merging clusters in the
A3558 chain
clearly indicate that
the signature of cluster merging on the radio emitting properties
of cluster galaxies and of the clusters themselves is 
complex.

The results emerging from our detailed statistical analysis of the
radio properties in connection with the dynamical state of the A3558 complex, 
can be summarised as follows:

\noindent
{\it (1)} the radio source counts in the A3558 complex
are consistent with the background source counts at a confidence level
of 99.6\%, despite the much higher optical density in this
region compared to the background;

\noindent
{\it (2)} the ratio between the radio and optical flux for the 
Shapley radio galaxies is not affected by the local galaxy density;

\noindent
{\it (3)} steep spectrum radio galaxies are not segregated
in a density-$\alpha_{13}^{22}$ diagram;
on the contrary,
galaxies with steep spectra ($\alpha_{13}^{22} \ge 1$) are preferentially
located at the periphery of the A3558 complex, in regions
where the galaxy density is much lower;

\noindent
{\it (4)} for radio powers logP$_{1.4}$ \gtsim $~$22.5 the radio-optical 
luminosity function 
for early type galaxies located in the A3558 region
is significantly lower than that derived by other authors (LO96 and references 
therein) for early type galaxies in clusters and in the field.

The implications of these results are extremely important.

Point {\it (2)} indicates that the very high
optical density of the merging environment in the Shapley
concentration has no effect on the radio galaxy emissivity.
In other words, given an optical magnitude, the radio power
of the associated radio source does not depend on the 
local environment. This is not entirely surprising if we think
that low power radio galaxies are almost only found in rich
clusters of galaxies. However, our analysis suggests that
not only the local density but also the merging environment
plays no role in the radio power distribution of radio sources.

Point {\it(1)} suggests that 
the dominant source population in our survey is 
the background, and that the much higher optical density in the Shapley
Concentration core and the ongoing merger is not reflected in a 
higher density of radio sources. 
In other words the Shapley Concentration would
have not been spotted using radio source counts.

The implication of point {\it (4)} is even more extreme, since
it suggests that the probability of a galaxy becoming a radio
source with logP \gtsim $~$22.5 is {\it lower} in the Shapley Concentration 
than in any other type of environment.

Our results can be explained in two different ways. In particular,

\noindent
{\it (a)} merging neither influences the probability of a 
galaxy to becoming a radio source nor increases its emissivity;

\noindent
{\it (b)} merging anticorrelates with radio emission, possibly
switching off previously existing radio galaxies.

\noindent
We point out that a preliminary analysis of the statistical radio 
properties in the merging complex formed by the three clusters
A3528, A3530 and A3532 give similar results (Venturi et
al. 1999).

Our results differ from those of Owen et al. (1999), who compared
the radio-optical properties of a merging and a relaxed
cluster, both at z$\sim$ 0.25, and concluded that cluster mergers
triggers radio emission, both in the form of nuclear emission and of
radio emission from starbursts.
The difference in our results suggests that the role of mergers on 
the radio emission properties of clusters is complex, and a variety 
of parameters, such as for example stage of merger, impact velocities
and timescales, are likely to play a role. 

The complexity of the situation also emerges from the fact 
that despite the above mentioned statistical results, some 
specific radio properties
in the A3558 chain seem to be the signature of cluster or group
merger, most remarkably the candidate radio halo at the centre
of A3562 and the extended emission around the radio galaxy J1332$-$3146. 
A detailed optical study of the spectral properties of
the galaxies in the A3558 chain 
indicates that this region is populated by a very large fraction of
blue galaxies. In particular $\sim$ 45\% of the galaxies in
this region show a blue excess, considered to be indicative of
star formation induced by mergers (Bardelli et al. 1998). 
According to the numerical simulations of
mergers in the A3558 complex, a shock front is expected in the
region between SC1329$-$313 and A3562 (Roettiger, Burns \& Loken 1993).

Further observational support for an ongoing merger
in this region of the A3558 complex comes from  
an X--ray spectral analysis carried out for the 
A3558 complex. ASCA observations by Hanami et al. (1999)
show that SC1329$-$313 exhibits remarkable peculiarities in this
energy band. In particular, its observed gas temperature is higher than
derived by the $\sigma$-T$_X$ relation and there is evidence that
the gas is presently in a recombination phase rather than in 
ionization equilibrium. According to Hanami et al. (1999) these
anomalies could be explained if SC1329$-$313 is an ongoing merger
(or if it just experienced a major merger event), responsible for
gas outflow from the core of the group.

\medskip

\vskip 1.0truecm
\noindent

{\bf Acknowledgements}

We thank Prof. R. Fanti for careful reading of the manuscripts
and for insightful comments. 

\noindent

T.V. acknowledges the receipt of a grant from CNR/CSIRO (Prot. N. 088864).

The Australia Telescope Compact Array is operated by the CSIRO Australia
Telescope National Facility.

%
%%%
%
%
{}

\end{document}